\documentclass[aip,jcp,amsmath,amssymb,floatfix,reprint,doi=true,citeautoscript]{revtex4-1}

\usepackage[utf8]{inputenc}
\usepackage{graphicx}
\usepackage{xcolor}
\usepackage{wrapfig}
\usepackage{bibunits}
\normalsize
\usepackage{booktabs}
\usepackage{bm}
\usepackage[colorlinks,allcolors=black,citecolor=blue,urlcolor=blue]{hyperref}

\graphicspath{{figures/}}

\defaultbibliography{paper,paper-control}
\defaultbibliographystyle{aipnum4-1}

\newcommand*\dd{\mathop{}\!\mathrm{d}}

\begin{document}

\def\mytitle{Explaining Principles of Tip-Enhanced Raman Images with Ab Initio Modeling}
\title{\mytitle}

\author{Krystof Brezina}
\affiliation{
Max Planck Institute for the Structure and Dynamics of Matter, Luruper Chaussee 149, 22761 Hamburg, Germany
}

\author{Yair Litman}
\affiliation{Max
Planck Institute for Polymer Research, Ackermannweg 10, 55128 Mainz,
Germany}
\affiliation{
Yusuf Hamied Department of Chemistry, University of Cambridge, Lensfield Road, Cambridge, CB2 1EW, United Kingdom
}

\author{Mariana Rossi}
\email{mariana.rossi@mpsd.mpg.de}
\affiliation{
Max Planck Institute for the Structure and Dynamics of Matter, Luruper Chaussee 149, 22761 Hamburg, Germany
}

\date{\today}

\begin{abstract}

Tip-enhanced Raman spectroscopy (TERS) is a powerful method for imaging vibrational motion and chemically characterizing surface-bound systems.
Theoretical simulations of TERS images often consider systems in isolation, ignoring any substrate support, such as metallic surfaces.
Here, we show that this omission leads to deviations from experimentally measured data through simulations with a new finite-field formulation of first-principles simulation of TERS spectra that can address extended, periodic systems. 
We show that TERS images of tetracyanoethylene on Ag(100) and defective MoS$_2$ monolayers calculated using isolated molecules or cluster models are qualitatively different from those calculated when accounting for the periodicity of the substrate. 
For Mg(II)-porphine on Ag(100), a system for which a direct experimental comparison is possible, these simulations prove to be crucial for explaining the spatial variation of TERS intensity patterns and allow us to uncover fundamental principles of TERS spectroscopy. 
We explain how and why surface interactions affect images of out-of-plane vibrational modes much more than those of in-plane modes, providing an important tool for the future interpretation of these images in more complex systems.

\noindent\textit{Keywords: Tip-enhanced Raman spectroscopy, normal-mode imaging, surface systems, first-principles calculations, density-functional theory}

\end{abstract}

{\maketitle}

\begin{bibunit}


Raman spectroscopy is a well established tool to elucidate the atomic structure, the atomic composition and the vibrational motion of matter in the gas phase or the condensed phase. 
Obtaining sufficient intensity of the scattered signal relies on the existence of a large amount of molecules or of the material~\cite{Craig1984}. 
Moreover, due to the diffraction limit, it is impossible to obtain signals from spatially resolved regions in the sub-nanometer range with typical visible light wavelengths. 
These properties prevent the acquisition of signals coming from single-molecules, defects or impurities at surfaces.

\begin{figure}[tb!]
    \centering
    \includegraphics[width=\linewidth]{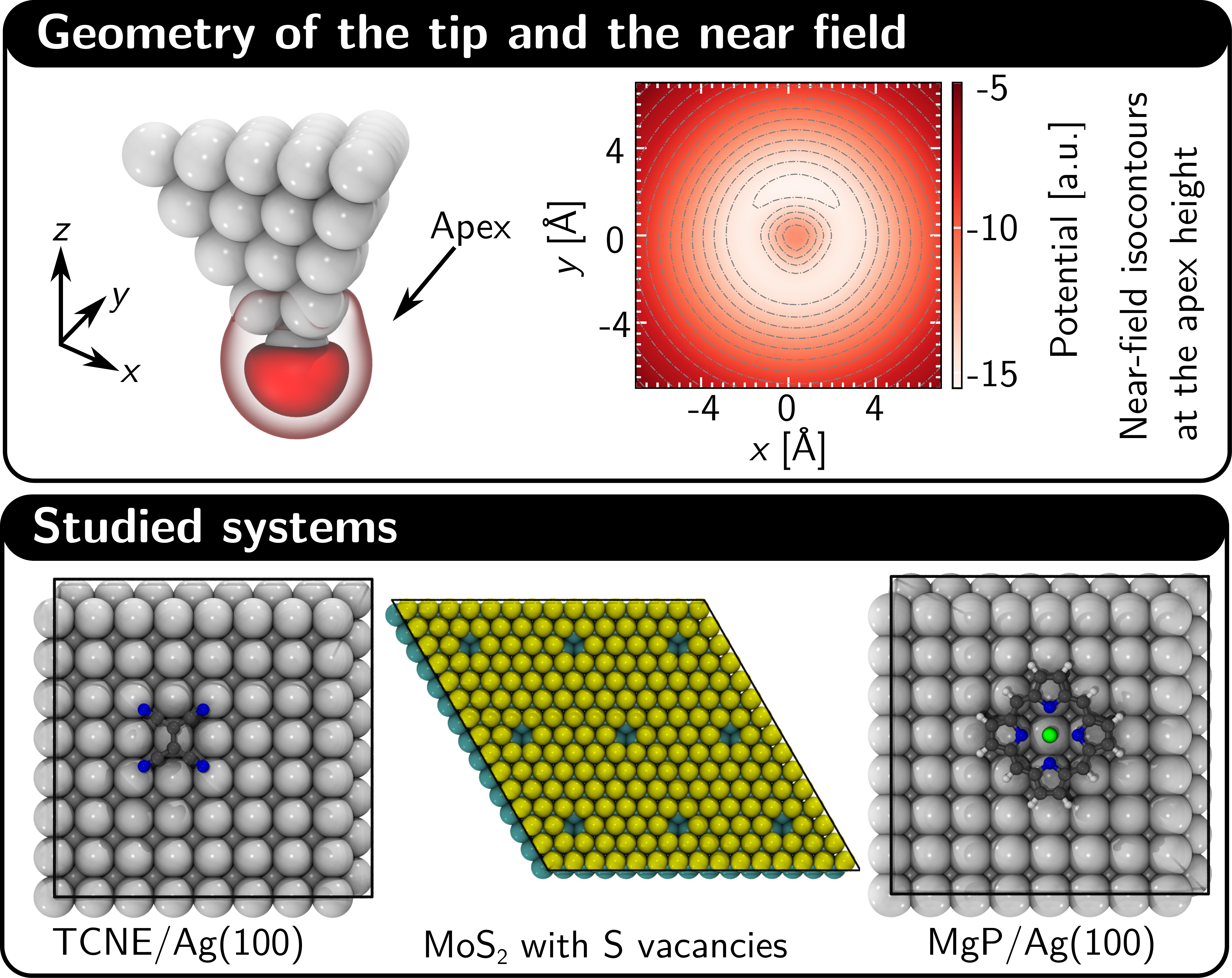}
    \caption{
    Top row (left to right): the geometry of the Ag tip including the spatial distribution of the atomistic near field (in red; darker shades correspond to a higher a intensity contour) and a horizontal cut through the potential at the level of the apex that illustrates the deviation of the near field from central symmetry.
    Bottom row (left to right): a birds-eye view of the surface systems studied in this work including tetracyanoethylene (TCNE)/Ag(100), an MoS$_2$ monolayer with 4\% sulfur vacancy concentration and magnesium(II) porphine (MgP)/Ag(100).
    All systems are treated under Born--von-Kármán boundary conditions and the boundary of the unit cell in each system is shown in black.
    }
    \label{fig:systems}
\end{figure}

Such drawbacks can be overcome with tip-enhanced Raman spectroscopy~\cite{Pozzi2017/10.1021/ACS.CHEMREV.6B00343, Shao2019/10.1007/S00216-018-1392-0, Cai2023/10.31635/CCSCHEM.022.202202287} (TERS).
This method borrows concepts from surface-enhanced Raman spectroscopy~\cite{Han2022/10.1038/s43586-021-00083-6} and relies primarily on the enhancement of the Raman signal by localized plasmonic resonances~\cite{Kneipp1999/10.1021/cr980133r}.
Specifically in the case of TERS, these are created by design at the junction formed between the metal surface and an atomically sharp metallic tip by the action of an external radiation~\cite{Zhang2016/10.1021/ACS.ANALCHEM.6B02093}.
This strong localization of the electromagnetic field at the nano-scale junction is the basis for the extreme spatial sensitivity of TERS that allows one to achieve sub-nanometer resolution and record Raman scattering from individual parts of molecules~\cite{Zrimsek2017, Steidtner2008/10.1103/PHYSREVLETT.100.236101, Zhang2019/10.1093/NSR/NWZ180, Lee2019/10.1038/s41586-019-1059-9}.

Alongside this electromagnetic enhancement mechanism due to plasmonic resonances, additional enhancement in the case of molecular adsorbates stems from the chemical interaction with the underlying conductive surface: this is known as the chemical enhancement.~\cite{Jensen2008/10.1039/B706023H, Gieseking2017/10.1039/c7fd00122c}
The scattered signal therefore contains contributions from the supporting surface to varying degrees.
On the one hand, these contributions can be exploited to study the specific properties of the interface and of surface--molecule interactions, but on the other hand, they can obfuscate certain molecular features.
Remarkably, TERS has been used to study the vibrational properties and imaging of single molecules including porphine derivatives~\cite{Lee2019/10.1038/s41586-019-1059-9, Zhang2019/10.1093/NSR/NWZ180} and shows great promise in applications involving, for example, DNA sequencing~\cite{Han2024/10.1021/JACS.4C12393}, molecular self-assembly~\cite{Zhang2022/10.1063/5.0091353/2841304}, defect characterization in van-der-Waals materials~\cite{Kato2019/10.1063/1.5080255, Jorio2024/10.1088/2053-1583/ad42ad, Akkoush2024/10.1002/PSSA.202300180} and surface-catalyzed chemical reactions~\cite{Bhattarai2019/10.1021/ACS.JPCLETT.9B00935}.

Theoretical simulations of vibrational spectra represent an invaluable complement to their experimental measurement as a predictive and interpretative means~\cite{Tuckerman2010}. 
While for standard (\textit{i.e.}, far-field) Raman spectroscopy, the simulation approaches are well established~\cite{Marx2009,Wan2013/10.1021/CT4005307,Marsalek2017/10.1021/ACS.JPCLETT.7B00391,Shang2018/10.1088/1367-2630/AACE6D}, for TERS the situation becomes more complex owing to the presence of the highly non-homogeneous plasmonic near field. 
Appropriate simulation approaches are a matter of active development~\cite{Jensen2008/10.1039/B706023H,Payton2014/10.1021/AR400075R, Liu2017/10.1021/ACSNANO.7B02058, Lee2019/10.1038/s41586-019-1059-9,Zhang2019/10.1093/NSR/NWZ180}. 
Existing methodologies have been generally successful in capturing the correct symmetries and shapes of the TERS images~\cite{Lee2019/10.1038/s41586-019-1059-9,Zhang2019/10.1093/NSR/NWZ180, Liu2017/10.1021/ACSNANO.7B02058}, however, an analysis of the literature shows that a method that obtains qualitative agreement with a broader range of experimental data is still lacking. 
For instance, Lee \textit{et al.} in their pioneering TERS study of Co(II)-tetraphenylporphyrin~\cite{Lee2019/10.1038/s41586-019-1059-9} faithfully matched the experimental 1156 cm$^{-1}$ asymmetric hydrogen bending mode image, but had qualitatively less success with the ``Maltese-cross'' mode image at 730 cm$^{-1}$, both simulated with a methodology that considered the isolated molecule in the gas phase.
In their discussion, the authors suggest the chemical interaction with the surface and the electronic screening as the primary culprits.

Motivated by these observations, we show in this paper that such shortcomings can be rectified by including the metal surface and the system's periodicity from first principles.
To this end, we develop a simulation methodology inspired by previous work of some of us~\cite{Litman2023/10.1021/acs.jpclett.3c01216}, which can fully account for both atomistic description of the plasmonic near fields (see the overview in Figure 1) and extended metallic surfaces and periodic systems in general within first-principles density-functional theory (DFT) simulations.
The technical details of the employed methodology can be found in Methods and in the Supporting Information (SI). 
We then employ this methodology to study TERS signals from defects and adsorbates as listed in Figure~\ref{fig:systems} and discussed in detail below.


\begin{figure*}[tb!]
    \centering
    \includegraphics[width=0.75\linewidth]{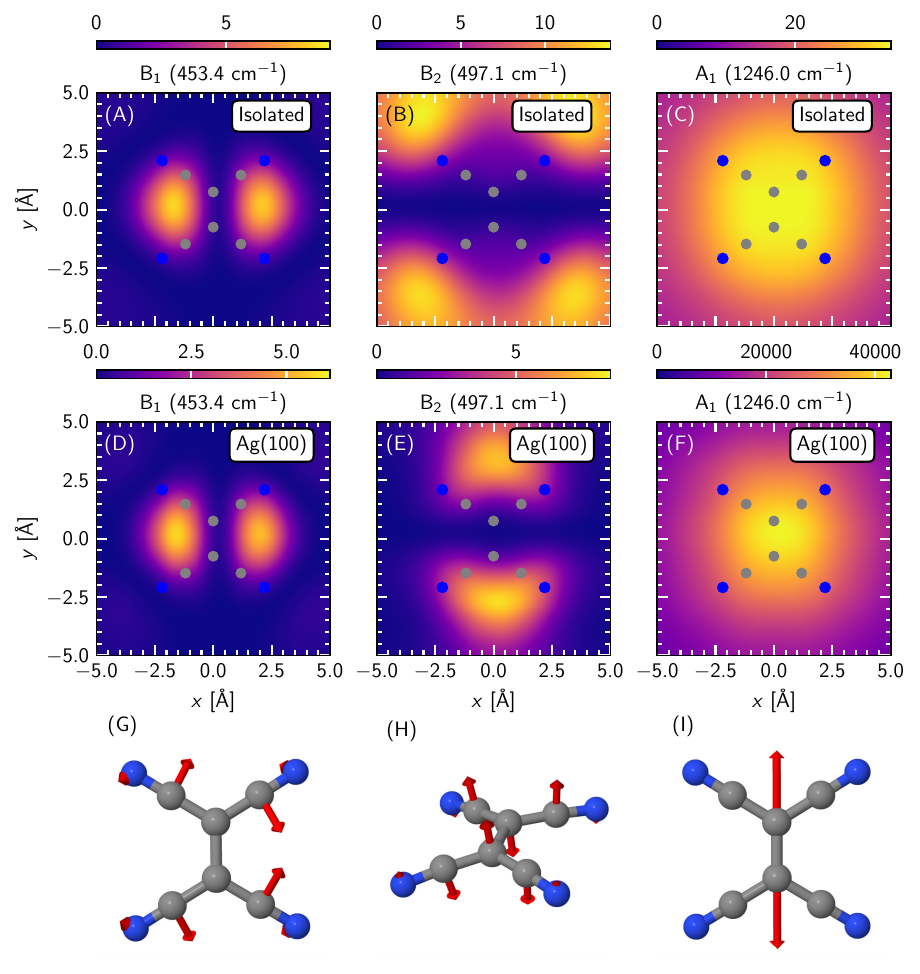}
    \caption{
    Simulation of TERS images of selected vibrational modes of TCNE/Ag(100).
    Panels A--C: Images obtained in the gas phase, however, using displacements calculated on the silver surface.
    Panels D--F: Images obtained on the explicit silver surface.
    In all panels A--F the corresponding color bars show the quantity $[\mathrm{d}\alpha_{zz} / \mathrm{d}Q(\omega)]^2 \cdot 10^4$ in the units of $\mathrm{e}^2 \mathrm{\AA}^2 \mathrm{V}^{-2}$ that is proportional to Raman intensity.
    Panels G--I: Snapshots of the corresponding vibrational modes of TCNE/Ag(100). 
    The silver surface was removed for clarity and the red arrow show the atomic components of the Cartesian normal mode vectors.
    In all panels the following atom color-coding applies: C --- gray, N --- blue.
    }
    \label{fig:TCNE}
\end{figure*}

\textbf{Metal surfaces are poorly approximated by small cluster models:} We benchmark our methodology by revisiting  tetracyanoethylene (TCNE) on Ag(100), a system that shows substrate effects on TERS images~\cite{Litman2023/10.1021/acs.jpclett.3c01216}. 
The simulations presented in Reference~\citenum{Litman2023/10.1021/acs.jpclett.3c01216} were obtained by approximating the Ag(100) surface by an Ag cluster. The images are shown for comparison in Figure~\ref{si-fig:original-tcne}, Section~\ref{si-sec:old-ters} of the SI. 

The TCNE molecule is fully planar in the gas phase, but bends its nitrogen atoms towards the surface upon adsorption, leading to a $C_\mathrm{2v}$ symmetric structure that is not disrupted by the surface when the molecule is at its optimal top adsorption site.
Relying on the current simulation methodology described in Section~\ref{si-sec:methodology} of the SI, we simulated TERS images for the same modes as in the previous work, but considering the periodic system with the Ag(100) surface. We considered a B$_1$ mode at 453.4~cm$^{-1}$ (Figure~\ref{fig:TCNE}, panels A, D and G), a B$_2$ mode at 497.1~cm$^{-1}$ (Figure~\ref{fig:TCNE}, panels B, E and H) and an A$_1$ mode at 1246.0~cm$^{-1}$ (Figure~\ref{fig:TCNE}, panels C, F and I).
In order to allow a quantitative estimate of the effect of the surface, we used the vibrational modes calculated for the surface-bound system to create atomic displacements in the gas phase.

Both the A$_1$ and B$_2$ mode images appear to be strongly affected by the Ag(100) surface.
The gas-phase TERS image of the A$_1$ mode is characterized by a single, broad peak positioned at the central C--C bond and spanning the whole TCNE molecule.
Upon adsorption, the shape of the image is preserved, but the surface causes a significant enhancement of the Raman intensity by a factor of $\sim10^3$.
In the case of the B$_2$ mode image, the Raman intensity remains similar upon adsorption to Ag(100), but it is the shape of the image that changes noticeably.
Specifically in the gas phase, the image exhibits a four-lobe pattern with peaks around the N atoms, whereas on the Ag(100) surface, the pattern is  distinctly two-lobe with peaks more closely localized to the region of the C atoms that mainly partake in the vibration.
Both findings are consistent with the previous results based on Ag clusters (\textit{cf.} Section~\ref{si-sec:old-ters} of the SI).
At variance, another result presented in Reference~\citenum{Litman2023/10.1021/acs.jpclett.3c01216} was an  alteration of the TERS image of the B$_1$ mode between the gas phase and the Ag(100) cluster. 
This is not reproduced on the periodic Ag(100) surface, where the isolated and surface-bound systems yield a qualitatively similar image. 
Hence, we attribute the previously observed effect in the B$_1$ mode to artifacts due to the finite size of the substrate and the asymmetry of the cluster.

The presented TCNE results illustrates two important points.
First, the surface can have a profound effect on the TERS image and it can manifest both in its intensity and its shape.
These effects are clearly not uniform across the different vibrational modes and the specific outcome depends on the vibrational pattern of each mode and the interaction with the underlying substrate.
Second, using finite clusters as an approximation to an extended, periodic metallic surface is not advisable: a certain cluster model might work well for some vibrations, but the same model can lead to pronounced artifacts in others.

\begin{figure}[tb!]
    \centering
    \includegraphics[width=0.75\linewidth]{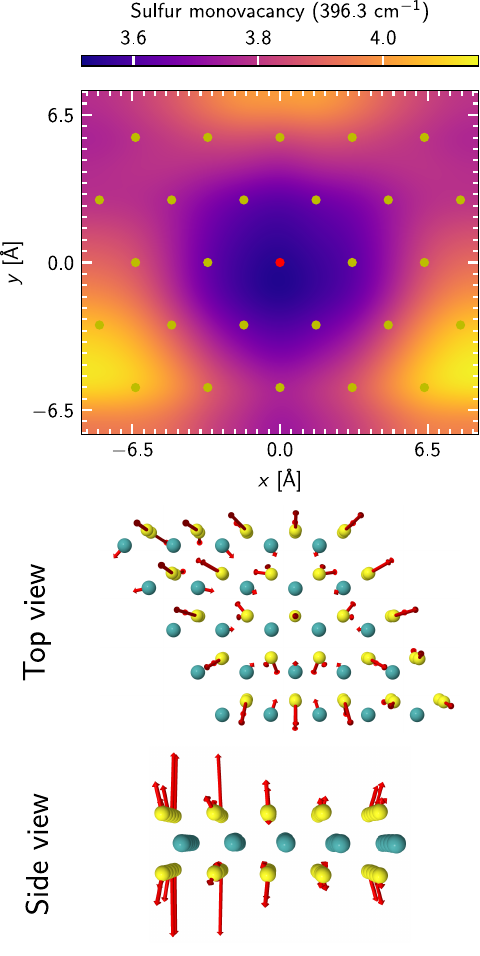}
    \caption{ 
        TERS imaging of the Raman-active defect-related equivalent of the A$_\mathrm{1g}$ vibration in an MoS$_2$ monolayer containing a sulfur monovacancy.
        The top panel shows the calculated TERS image. 
        The positions of the top-layer sulfur atoms are given by the yellow circles and the location of the defect at (0, 0) $\mathrm{\AA}$ is marked by the red dot.
        The TERS intensity is shown in the units of 10$^1$ $\mathrm{e}^2 \ \mathrm{\AA}^2 \ \mathrm{V}^{-2}$.
        The snapshots show a top view and a side view of the vibration on a 5$\times$5 unit cell.
        In these snapshots, the sulfur atoms are shown in yellow, the molybdenum atoms in turquoise and the cartesian atomic components of the normal mode vector as red arrows.
    }
    \label{fig:MoS2}
\end{figure}

\textbf{Vibrational fingerprints of defects in 2D systems can be long-ranged:} As a second example of the importance of the inclusion of periodicity in the simulation of TERS images, we revisit the investigation of defect-related vibrational modes in monolayer MoS$_2$. 
In particular, we look for vibrational modes stemming from a sulfur monovacancy, which is the most abundant defect under usual experimental conditions~\cite{Li2022/10.1039/d2na00636g, Ascrizzi2025/10.1021/acs.jpcc.4c08631}. 
These calculations were previously attempted in Reference~\citenum{Akkoush2024/10.1002/PSSA.202300180} using a MoS$_2$ flake instead of the periodic system. 
The results suggested a drop in TERS intensity around the vacancy site, but a clear interpretation is obfuscated by a large spatial variation of the TERS intensity across the flake due to the finite size and asymmetry of the system.
We employ the proposed simulation approach for TERS imaging of the A$_\mathrm{1g}$ vibration in periodic monolayers, which produces a strong signal in TERS  due to its out-of-plane character.

As shown in Section~\ref{si-sec:pristine-mos2} of the SI, in the pristine monolayer, because of the perfectly concerted motion of the S atoms, the simulated image mirrors the geometry of this vibration and exhibits a very uniform intensity across the system. 
In fact, the minute intensity variations around the S atoms serve as a probe of the accuracy of our simulations, which we prove to be at least on the order of 10$^{-4}$~$\mathrm{e}^2 \ \mathrm{\AA}^2 \ \mathrm{V}^{-2}$. This accuracy is only achievable with tight electronic-structure convergence criteria (see Section~\ref{si-sec:methodology} for details).

The presence of defects lowers the symmetry of the system and effectively folds different phonon branches. 
As a consequence, multiple $\Gamma$-point modes appear within the smaller Brillouin zone in the wavenumber range corresponding to the pristine A$_\mathrm{1g}$ vibrational band.
Here, we show in Figure~\ref{fig:MoS2} the TERS image of the mode that corresponds to the highest Raman intensity in this wavenumber range as calculated for a 5$\times$5 unit cell (see Section~\ref{si-sec:methodology} of the SI for details).
As depicted in the same Figure, this vibration exhibits a $C_3$ symmetry axis centered at the vacancy and maintains the general out-of-plane character of the motion of the sulfur atoms~\cite{Zhao2019/10.48550/arxiv.1904.09789}.
However, the motion becomes more complex as the vibrational direction of the S atoms is no longer collinear and certain Mo atoms also partake in it when the vacancy is present.
The corresponding TERS image features a pattern that appears as a consequence of the defect: a lower Raman intensity region centered at the defect site surrounded by a ring of a higher intensity that clearly shows a $C_3$ symmetry.
Obtaining such symmetry is only possible in a periodic system, or would necessitate flakes much larger than the characteristic length-scale of the spectral signal and that would not break the three-fold symmetry of this system. 
That would render the simulation prohibitive, as the effect of the vacancy on the TERS image extends its range beyond 1~nm.
While the experimental TERS imaging of vacancy defects in MoS$_2$ at a sub-nanometer resolution remains a challenge~\cite{Jorio2024/10.1088/2053-1583/ad42ad}, a simulation of an extended periodic system sets a reasonable starting point for a potential interpretation of such experiments.

\begin{figure*}[tb!]
    \centering
    \includegraphics[width=0.75\linewidth]{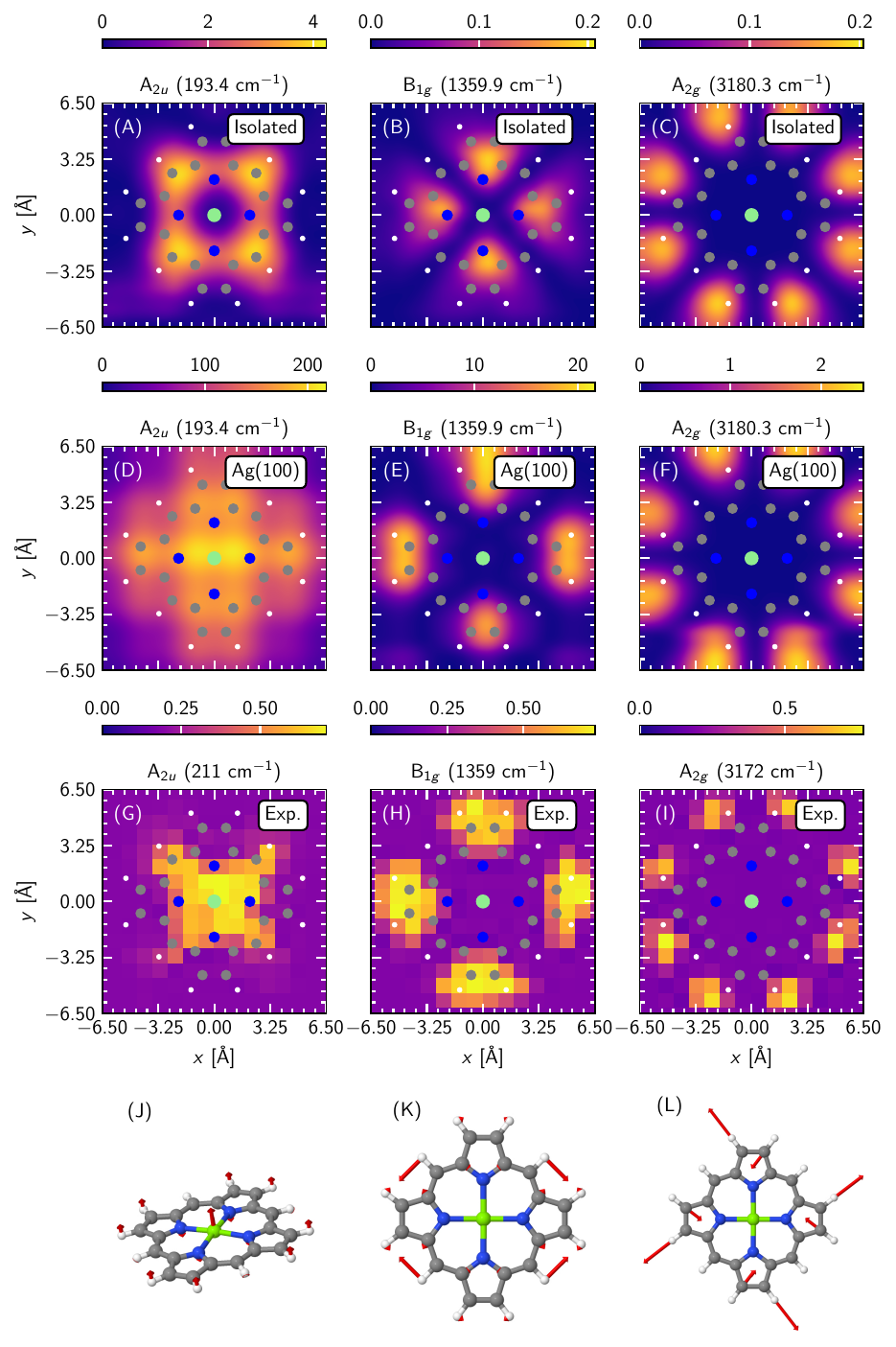}
    \caption{
        Simulation of TERS images of selected vibrational modes of MgP/Ag(100).
        Panels A--C: Images obtained in the gas phase, however, using displacements calculated on the silver surface.
        Panels D--F: Images obtained on the explicit silver surface.
        In all panels A--F the corresponding color bars show $[\mathrm{d}\alpha_{zz} / \mathrm{d}Q(\omega)]^2 \cdot 10^4$ in the units of $\mathrm{e}^2 \mathrm{\AA}^2 \mathrm{V}^{-2}$, which is proportional to Raman intensity.
        Panels G--I: Snapshots of the corresponding vibrational modes of MgP/Ag(100).
        The silver surface was removed for clarity and the red arrows show the atomic components of the Cartesian normal mode vectors.
        In all panels the following color-coding of atomic species applies: Mg green, C gray, N blue, H white.
    }
    \label{fig:MgP}
\end{figure*}

\textbf{Surface screening qualitatively changes TERS patterns:} Finally, we present our findings concerning the simulation of TERS images of the MgP/Ag(100) system, which leads to a new understanding of the principles underlying TERS image patterns. 
For this system, there is directly comparable experimental data available~\cite{Zhang2019/10.1093/NSR/NWZ180}.
Upon adsorption on the Ag(100) substrate, the MgP molecule only undergoes a negligible deviation from its gas-phase planar $D_\mathrm{4h}$ symmetry and adopts a position where the central Mg ion sits on top of an Ag atom in the surface layer of the metal slab.
We inspect three representative normal modes across the frequency spectrum: a 193.4~cm$^{-1}$ A$_\mathrm{2u}$ mode dominated by the out-of-plane vibration of the Mg atom (Figure~\ref{fig:MgP}, panel J), a 1359.9~cm$^{-1}$ B$_\mathrm{1g}$ mode which captures the asymmetric breathing of the pyrrole rings (Figure~\ref{fig:MgP}, panel K) and, finally, a 3180.3~cm$^{-1}$ A$_\mathrm{2g}$ asymmetric hydrogen stretching mode (Figure~\ref{fig:MgP}, panel L).
All of these modes exhibit a rich spatial variation of the TERS signal as demonstrated by the experimental images in panels G--I of Figure~\ref{fig:MgP}, which were originally published by Zhang \textit{et al.} in  Reference~\citenum{Zhang2019/10.1093/NSR/NWZ180}.
Specifically, the A$_\mathrm{2u}$ mode shows the highest intensity above the central Mg atom with weak tails reaching towards the bridge CH groups; the B$_\mathrm{1g}$ mode has a distinct four-peak structure with maxima around the pairs of distal carbon atoms of the pyrrole rings;
the A$_\mathrm{2g}$ mode shows an 8-peak pattern located on the pyrrole hydrogen atoms.

An important point to highlight before we discuss the TERS images in more detail is that all of the images on the surface (for example, the B$_2$ mode of TCNE/Ag(100) in panel E of Figure~\ref{fig:TCNE}) are affected to a varying degree by asymmetry.
When there is no reason for such asymmetry to be present due to the geometry of the scattering subsystem, such as in the case of the systems presented in this paper, the only source of remaining asymmetry is the atomistic description of the tip potential.
This potential is inherently asymmetric with respect to the symmetry of the molecular systems (both TCNE and MgP alike) since it bears an imprint of its original skewed trigonal-pyramidal silver cluster geometry~\cite{Litman2023/10.1021/acs.jpclett.3c01216} (see the horizontal sections of the atomistic near field used in our calculations in Figure~\ref{fig:systems}).
In our calculations, the tip is positioned such that it breaks the symmetry of the $y=0$ mirror plane; it is therefore fully consistent with the fact that all of our observed asymmetry manifests along the $y$-axis.
We have confirmed this hypothesis explicitly by fitting a dipolar potential corresponding to the $z$-direction to the full tip potential. 
This fit has a $C_\infty$ axis and does not break any underlying molecular symmetries.
As demonstrated in Section~\ref{si-sec:tip-symmetry} of the SI, using this dipolar field to recreate the image of the B$_\mathrm{1g}$ MgP/Ag(100) mode creates a fully symmetric image.
This finding opens an intriguing debate about how the tip geometry translates into the (a)symmetry of experimental images as the atomistic shape of the tip is typically not known.
In turn, one could  gain the ability to infer the shape of the tip by inspecting the shapes of the measured image after having first mapped out the relationship between the two computationally.

In order to understand the origin of the TERS patterns of MgP/Ag(100), we first calculate the TERS images of MgP molecule in the gas phase and show the obtained results in panels A--C of Figure~\ref{fig:MgP}.
It is clear that in certain cases, the gas-phase approximation can give rise to images that are very close to the experiment, such as the A$_\mathrm{2g}$ mode displayed in panel C.
However, this is not true in general.
The 4-peak image of the B$_\mathrm{1g}$ mode (panel B) is described correctly only from a  qualitative perspective but the peak positions are incorrect: the gas-phase image shows peak maxima around the pyrrole nitrogen atoms, whereas the experiment has peaks located around the distal pyrrole carbon atoms.
The gas-phase simulation of the A$_\mathrm{2u}$ mode (panel A) is incorrect even from a qualitative viewpoint, as the image shows a pronounced minimum at the location of the central Mg atom where the experiment shows the highest intensity.
Once again, we thus find that the gas-phase approximation is inaccurate as a means of comparison to surface-bound experimental data and shows alterations that depend on the nature of the specific vibrational mode in question and cannot be known \textit{a priori}.

Now we turn our attention to the images simulated with the inclusion of an atomistic Ag(100) periodic surface as shown in panels D--F of Figure~\ref{fig:MgP}.
These results show a significantly improved agreement with the experiment in all studied modes.
The A$_\mathrm{2g}$ mode (panel F), which was already well described in the gas phase, retains this quality on the surface.
The peaks of the B$_\mathrm{1g}$ mode image (panel E) are now positioned consistently with the experimental data.
Finally and perhaps most interestingly, the surface simulation (panel D) reproduces the region of high intensity in the middle of the A$_\mathrm{2u}$ mode image observed experimentally (panel G), in contrast to the gas-phase simulation (panel A). Therefore, it allows us to link this intensity with the surface--molecule interaction and illustrates the active involvement of the metal substrate in the shaping of TERS images. At the same time that this simulation captures the build up of TERS intensity above the Mg atom, it also fails to reproduce the tails of the experimental peaks extending towards the bridge CH groups. These two observations deserve further discussion.

\begin{figure}[tb!]
    \centering
    \includegraphics[width=\linewidth]{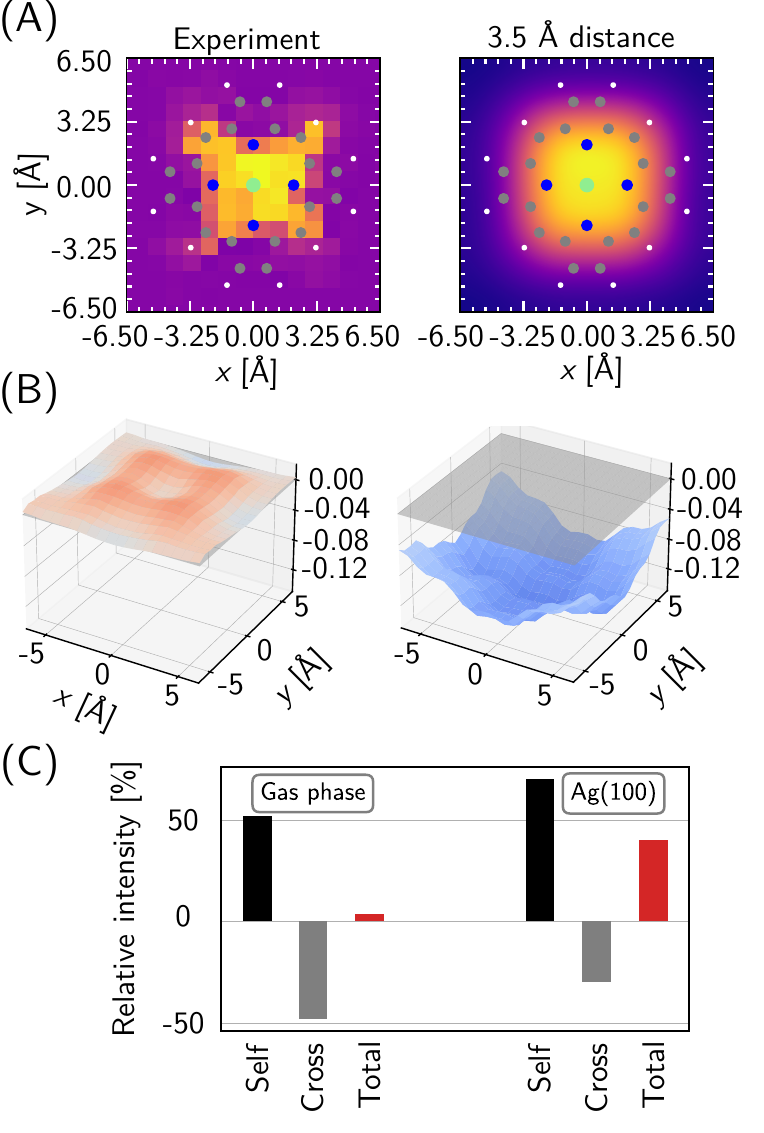}
    \caption{
    Further analysis of the shape and intensity patterns in the A$_\mathrm{2u}$ mode of MgP/Ag(100).
    Panel A: The left plot shows the experimental image for reference (same data as Figure~\ref{fig:MgP}G). 
    The right plot shows an image calculated with the surface--molecule distance empirically increased to 3.5~\AA, which reproduces the intensity tails over the bridge C-H groups observed in the experiment.
    Panel B: The TERS amplitude calculated in the gas phase (left) and on the surface (right) shown as a 3D surface.
    The gray plane shows the zero magnitude.
    The employed color map ranges from negative (darker blue) to positive (darker red) with the point of zero value in white.
    Panel C: The decomposition of the TERS intensity into self and cross terms shows the near complete cancellation of the terms in the gas phase and and incomplete cancellation on the surface. 
    }
    \label{fig:decomposition}
\end{figure}

We start by explaining why the shape of the image in Fig.~\ref{fig:MgP} panel D does not match more faithfully the shape of the experimental image reported in panel G.  
We find that the shape of the resulting image for this mode depends on the equilibrium surface--molecule separation distance. Computationally, this distance depends strongly on the choice of the DFT functional and, in particular, of the employed dispersion correction as we show in Section~\ref{si-sec:distance} of the SI. 
Indeed, we find that screened many-body van-der-Waals corrections~\cite{Tkatchenko2012/10.1103/physrevlett.108.236402} increase the molecule--surface distance by 0.13~\AA\ (see Table~\ref{si-tab:distances} in Section~\ref{si-sec:distance} of the SI). 
Taking this into consideration, we empirically varied the surface--molecule distance and found that by placing the molecule at around 3.5~\AA\, from the surface yields the TERS pattern reported in Figure~\ref{fig:decomposition}A (right panel). That pattern matches the experimental data, also reproduced in Figure~\ref{fig:decomposition}A (left panel), much more closely and suggests that this observation could be indirectly used to determine surface--molecule distances of single molecules, which is often hard to measure in experiment.

We then continue with an analysis of how the surface acts to dramatically change the TERS image of some modes (such as the A$_\mathrm{2u}$ mode of MgP) and to leave others practically unaltered.
As explained in the Methods section, under suitable approximations the TERS intensity can be calculated as
\begin{equation}
     I_{zz}(\omega_k, \mathbf{R}_\mathrm{tip})
     \propto
     \left[\frac{\partial \alpha_{zz}^\mathrm{local}(\mathbf{R}_\mathrm{tip})}{\partial Q_k}\right]^2
     \equiv
     A^2_{zz}(\omega_k, \mathbf{R}_\mathrm{tip})
\end{equation}
where $\alpha_{zz}^\mathrm{local}(\mathbf{R}_\mathrm{tip})$ is the $zz$-component of the tip-position $\mathbf{R}_\mathrm{tip}$ dependent local polarizability and $Q_k$ is the normal coordinate of a vibration with characteristic frequency $\omega_k$.
To gain further insight, we inspect the quantity $A_{zz}$, which describes how the polarizability changes with the vibration, carries a sign and, therefore, can be called the TERS amplitude.
This quantity is shown for the gas-phase and the surface-bound MgP A$_\mathrm{2u}$ vibration in Figure~\ref{fig:decomposition}B.
The gas-phase amplitude is almost exclusively positive and features the central minimum at the Mg atom surrounded by the four peaks over the bridge C atoms. 
In the surface-bound case, the general shape of the amplitude in the vicinity of the molecule remains the same: the Mg atom features a minimum and the bridge carbons four regions of higher intensity.
However, the amplitude is negative, which leads to the emergence of the central maximum in the corresponding TERS image. 

The variations of the amplitude around the molecule are a non-trivial result of the specific interaction of the vibrating molecule with the plasmonic near field and the underlying metal surface.
The fundamental cause of the relative sign change in $A_{zz}$, however, does not depend on the details of the near-field. 
We find that the primary source of this effect is the screening of the induced dipole by the Ag(100) substrate as the molecular vibration takes place.
We explicitly demonstrate this by tracking the changes in electron density response as a function of vibrational displacement and provide additional discussion in Section~\ref{si-sec:density-changes} of the SI.

Predicting the quantitative scale of the change in $A_{zz}$ requires first-principles modeling as it depends on the specific chemistry of the system.
However, we can infer from the presented $A_{zz}$ plots in Figure~\ref{fig:decomposition}B that the sign change is consistent throughout the whole lateral extent of the image, and thus points to an effect that is independent of the distribution of the near-field. 
As such, it would also be noticeable in standard Raman intensity, and this could be used as a useful qualitative marker to predict the presence of such screening effects. This effect can be understood as following: In the gas-phase, the molecular distortion along this normal mode leads to an increase in the corresponding polarizability tensor component, meaning that response of the electronic density is enhanced when the molecule is distorted. When the molecule is adsorbed, instead, the electrons of the surface act to screen this effect and instead lead to a decrease in the electronic density response when the molecule is displaced along the same normal mode. The near field only enhances this effect.

In essence, only modes that have a non-zero Raman intensity along the scattering direction can exhibit this sign-change effect.
This is the case for the A$_\mathrm{2u}$ (out-of-plane) mode in MgP/Ag(100).
The remaining two in-plane modes shown in Figure~\ref{fig:MgP} have vanishing Raman intensities along $z$ owing to symmetry and, therefore, these modes can only exhibit changes in their intensity enhancement patterns that do not change the symmetry nodal planes.
We note that this reasoning also explains why the TERS pattern in the B$_1$ mode of TCNE (Figure~\ref{fig:TCNE}A vs. \ref{fig:TCNE}D) on the Ag(100) surface is less altered than what was observed on an Ag(100) cluster where the symmetry is broken.
Finally, we note in passing that our calculations work with tip models of limited size, which translates into an arbitrary scaling of the near-field potential (see the Methods section).
We find that such scaling, within the limits of linear polarization, maps into a scaling of the overall TERS intensity, but does not affect the relative sign change in $A_{zz}$ between the gas-phase and the surface systems.

Finally, we attempt to gain a deeper understanding of local and non-local contributions to TERS intensity enhancement patterns. 
The intensity patterns are often qualitatively described as having a large magnitude where the atomic motion within a normal mode is large~\cite{Duan2016/10.1002/anie.201508218}. 
We check this assumption by performing a decomposition of the TERS intensity into atomic terms.
We can use a chain rule to write
\begin{equation}
\begin{split}
     A_{zz}(\omega_k, \mathbf{R}_\mathrm{tip})
     &=
     \frac{\partial \alpha_{zz}^\mathrm{local}(\mathbf{R}_\mathrm{tip})}{\partial Q_k} \\
     &=
     \sum_{i} \frac{\partial \alpha_{zz}^\mathrm{local}(\mathbf{R}_\mathrm{tip})}{\partial q_i} a_{k,i},
\end{split}
\label{eq:decomposition}
\end{equation}
where $\mathbf{a}_k$ is the Cartesian $k$-th normal mode vector and $q_i$ are Cartesian displacements.
Therefore, when calculating the TERS intensity, there will be terms that involve only displacements on the same atom and cross-terms that multiply displacements on different atoms
\begin{equation}
    I_{zz}(\omega_k, \mathbf{R}_\mathrm{tip})
    =
    I_{zz}^\mathrm{self}(\omega_k, \mathbf{R}_\mathrm{tip})
    +
    I_{zz}^\mathrm{cross}(\omega_k, \mathbf{R}_\mathrm{tip}).
\end{equation}
While the self terms are non-negative, the cross terms can assume negative values.

Such a decomposition can provide an atomistic perspective on the emergence of the observed sign-change effect as discussed in detail in Section~\ref{si-sec:atomic-decomposition} of the SI.
Both $I_{zz}^\mathrm{self}$ and $I_{zz}^\mathrm{cross} $ are shown for the tip positioned right above the Mg atom for the A$_\mathrm{2u}$ MgP mode in Figure~\ref{fig:decomposition}D.
In the gas phase, the $I_{zz}^\mathrm{cross}$ almost perfectly cancels the $I_{zz}^\mathrm{self}$, leading to the very small intensity observed at the origin despite the fact that the Mg atom has by far the largest amplitude of motion in this mode.
At variance, $I_{zz}^\mathrm{cross}$ on the surface has a noticeably smaller magnitude, which results in an incomplete cancellation of the two and, consequently, the observed non-zero intensity peak at the origin.
This perspective shines a new light on the interpretation of TERS images: the presence of cross terms connecting different atoms can lead to the emergence of non-trivial patterns beyond those reflecting the normal-mode geometry.
Further discussion and interpretation of this spectral decomposition is provided in Section~\ref{si-sec:locality-of-density-response} of the SI.


In conclusion, the results we have presented led to a much deeper understanding regarding the interpretation of TERS images from experiments and simulations.
The simulations we performed on TCNE/Ag(100) and defective MoS$_2$ monolayers proved that approximating the surface as a cluster (or omitting it altogether in the case of molecules in a so-called gas-phase calculation) is inaccurate, leading to images that contain  artifacts. 
A proper handling of periodicity is crucial for an accurate modeling.
The simulations and direct comparison to experiment of the MgP/Ag(100) system led to new insights into the origins of the shape of TERS images, including its dependence on the surface--molecule distance and, importantly, how it is impacted by the screening effects of the surface.
We find that such screening plays a fundamental role in the overall amplitude of the vibrational normal modes featuring out-of-plane changes of polarization (such as the A$_{2u}$ mode in MgP). 
Variations that lead to changes in the sign of the amplitude can completely alter the TERS images of these modes with respect to those obtained from isolated-molecule simulations. 
In addition, we provided a quantitative estimate of the non-local contributions to intensity patterns of TERS images, thus showing that the common assumption that TERS images have larger magnitudes where the atomic motion is larger is often invalid. 

This understanding was only achieved due to the methodology we developed. 
We presented a computationally efficient, first-principles framework for calculating TERS spectra and images on periodic substrates, using finite-field perturbations.
The method is implemented in the FHI-aims electronic structure software~\cite{Blum2009/10.1016/j.cpc.2009.06.022, abbott2025roadmap-b1e} and thus widely available to the community. 
Even though we have presented TERS intensity patterns obtained within a harmonic approximation for non-resonant Raman scattering, the methodology is very flexible and it is readily possible to couple it with dynamical methods that allow the study of anharmonic vibrational motion, of reactive events, and the inclusion of nuclear quantum effects.

We identify the main approximations of the current  methodology as the following: (1) Limitation to relatively small tip models for the plasmonic near-field distribution; (2) The use of static perturbations that preclude the treatment of resonant TERS signals;  (3) The assumption of non-overlapping electronic densities between tip and molecule or substrate, that precludes the treatment of the contact regime; (4) Lack of reradiation enhancement mechanisms~\cite{AusmanSchatz2009}.
Augmenting the method along any of these directions will increase its realm of application, sometimes to areas showing interesting new spectroscopic behavior~\cite{cirera2021charge-0eb}. 
While some of these extensions require new physics to be incorporated in the model, others, such as larger tip sizes and resonant responses, would benefit from modern machine-learning models capable of dealing with such complex electronic responses. 
It is, nevertheless, very refreshing to see that the method presented in this work is very well suited to model many current TERS experimental setups and unravel new understanding of these measurements for systems of essentially arbitrary complexity.

\section*{Methods}

We briefly outline the main underlying principles of the employed methodology in the following paragraph. 
A fully first-principles solution of TERS would call for a real-time propagation of the quantum system that includes both the scattering subsystem and the tip under the influence of the oscillating far field perturbation.
While this is technically possible, the high computational cost of explicitly propagating the electronic degrees of freedom limits the applicability to the smallest scattering subsystems~\cite{Liu2023/10.1021/acsnano.3c00261,Zhao2006/10.1021/NL0607378, Jensen2007/10.1021/JP067634Y}.
Therefore, most commonly used approaches rely on a combination of a phenomenological localization of the near field and either a polarizable force-field description of the surface~\cite{Payton2014/10.1021/AR400075R, Liu2017/10.1021/ACSNANO.7B02058}, or omit the surface altogether and simulate the molecules as isolated entities~\cite{Lee2019/10.1038/s41586-019-1059-9,Zhang2019/10.1093/NSR/NWZ180}.

With the aim to account for the substrate in efficient, first-principles TERS simulations, some of us~\cite{Litman2023/10.1021/acs.jpclett.3c01216} have formulated the following approximate theory.
In this framework, the time-dependent problem of the whole interacting system including the surface, the molecule (or any spatially-localized chemical environment) and the tip under the influence of an external periodic electromagnetic radiation is transformed into a  static problem with the perturbed Hamiltonian 
\begin{equation}
\begin{split}
    \hat{H}
    &=
    \hat{H}_0^{\mathrm{sc}}
    + 
    \hat{\Phi}_0(\mathbf{R}_{\mathrm{tip}}) \\
    & 
    + E_z
    \left\{
      -\hat{\mu}_z^{\mathrm{sc}} 
      + 
      \left[\frac{\partial \tilde{\hat{\Phi}}(\omega_\mathrm{p}; \mathbf{R}_{\mathrm{tip}})}{\partial E_z} 
    \right]_{E_z = 0}\right\}.
\end{split}
\label{eq:hamiltonian}
\end{equation}
In this expression, $\hat{H}_0^{\mathrm{(sc)}}$ represents the Hamiltonian terms pertaining to the scattering subsystem (\textit{e.g}., molecule and its supporting surface or defect center in a material), $E_z$ is the intensity of the component perpendicular to the surface plane of the incoming far field, $\hat{\mu}_z^{\mathrm{sc}}$ is the $z$-component of the dipole operator of the scattering subsystem, $\mathbf{R}_\mathrm{tip}$ is the position of the tip and $\hat{\Phi}_0$ is the unperturbed electrostatic scalar field of the tip. 
The key quantity $\tilde{\hat{\Phi}}(\omega_\mathrm{p})$ represents the Fourier component at the plasmon excitation frequency $\omega_\mathrm{p}$ of the 
oscillating electrostatic potential  $\hat{\Phi}(t)$ of the isolated tip under the influence of the incoming radiation.
The quantity $\hat{\Phi}_0$ can be neglected at a price of introducing a reasonably small error (as shown in Section~\ref{si-sec:groundstate} of the SI), leading to a significant computational speed up.

Importantly, the quantity $\hat{\Phi}(t)$ is calculated for an isolated tip using real-time TDDFT~\cite{Yabana1996/10.1103/physrevb.54.4484, Tancogne-Dejean2020/10.1063/1.5142502/954926} and the derivative with respect to $E_z$ is obtained numerically by performing simulations at several field strengths. This must only be done once for a given tip geometry~\cite{Litman2023/10.1021/acs.jpclett.3c01216} and the spatial distribution of the derivative of the resulting Fourier-transformed quantity $\tilde{\hat{\Phi}}(\omega_\mathrm{p})$ is stored and used off the shelf for arbitrary scattering subsystems.

Equation~\ref{eq:hamiltonian} is valid under three reasonable assumptions: (a) a tip distance large enough that the interaction between tip and molecule or defect is dominated by electrostatics, (b) $E_z$ weak enough that the induced tip polarization varies linearly with its intensity, (c) non-resonant Raman scattering so that only the static problem is relevant.
Note that we limit ourselves to the $z$-components of all vector quantities: these are experimentally relevant as most measurements are realized with the laser oriented approximately parallel to the surface normal direction (chosen to be $z$), with detection in the back-scattering regime~\cite{Zhang2016/10.1021/ACS.ANALCHEM.6B02093}.

Using $\hat{H}$ in Eq.~\ref{eq:hamiltonian}, density-functional perturbation theory~\cite{Baroni2001/10.1103/revmodphys.73.515, Shang2018/10.1088/1367-2630/AACE6D} (DFPT) can be applied in order to calculate the relevant $zz$-component of the spatially-dependent polarizabilty tensor $\alpha_{zz}(\mathbf{R}_\mathrm{tip})$. Under the harmonic approximation, the corresponding Raman intensity is given by  
\begin{equation}
    I_{zz}(\omega_k, \mathbf{R}_\mathrm{tip})
    \propto
    \left[\frac{\partial \alpha_{zz}^\mathrm{local}(\mathbf{R}_\mathrm{tip})}{\partial Q_k} \right]^2.
    \label{eq:raman}
\end{equation}
Here, $Q_k$ is the normal coordinate of the $k$-th vibrational mode with eigenfrequency $\omega_k$.

A drawback of this methodology is that the real-space DFPT formulation is not easily applicable to periodic systems under non-homogeneous electric perturbations, in particular if the surface is metallic~\cite{Shang2018/10.1088/1367-2630/AACE6D}. 
The present work reformulates this methodology for use in periodic systems, which is achieved by a replacement of the DFPT-based step with a finite-field calculation.
Specifically, through a self-consistent solution of $\hat{H}$ in Equation~\ref{eq:hamiltonian} under Born--von-Kármán boundary conditions, one obtains an electron density $\rho(\mathbf{r}; \mathbf{R}_\mathrm{tip})$ and, subsequently a $z$-component of a dipole moment via a real-space integration
\begin{equation}
    \mu^\mathrm{sc}_z(\mathbf{R}_\mathrm{tip})
    =
    \int_{\text{unit cell}} \mathrm{d}\mathbf{r} \ z \rho(\mathbf{r}; \mathbf{R}_\mathrm{tip}).
\end{equation}
This dipole component is always physically meaningful as the $z$-direction is (effectively) aperiodic in the slab geometry. A dipole-correction is always applied.
The next key step relies on the application of a finite homogeneous electric field of magnitude $\Delta E_z$ (which turns on the whole perturbation term including the near-field coupling in Equation~\ref{eq:hamiltonian}) to calculate the $zz$-component of the polarizability tensor through a finite difference as 
\begin{equation}
    \alpha_{zz}^\mathrm{local}(\mathbf{R}_\mathrm{tip})
    =
    \frac{\partial \mu_z^{\mathrm{sc}}(\mathbf{R}_\mathrm{tip})}{\partial E_z}
    =
    \frac{\Delta \mu_z^{\mathrm{sc}}(\mathbf{R}_\mathrm{tip})}{\Delta E_z};
    \label{eq:polarizability}
\end{equation}
since we are ensuring being in the regime of linear polarization, the finite difference is exact.
In the expression above, $\Delta \mu_z^{\mathrm{sc}}$ represents the induced dipole moment
\begin{equation}
\begin{split}
    \Delta \mu_z^{\mathrm{sc}}(\mathbf{R}_\mathrm{tip})
    &=
    \mu_z^{\mathrm{sc}}(\mathbf{R}_\mathrm{tip}, E_z=\Delta E_z) \\
    &- \mu_z^{\mathrm{sc}}( \mathbf{R}_\mathrm{tip}, E_z = 0).   
\end{split}
\end{equation}
Once the polarizabilities are known, one can proceed to the calculation of TERS images under the harmonic approximation using Equation~\ref{eq:raman}.
Note that under open boundary conditions, the two formulations are identical as we demonstrate in Section~\ref{si-sec:dfpt-vs-ff} of the SI.
We have implemented this new approach in FHI-aims and provide a thorough description of the specific computational details in Section~\ref{si-sec:methodology} of the SI.

\begin{acknowledgments}

K.B. thanks prof. Z. C. Dong and his group for sharing the original MgP/Ag(100) experimental TERS mapping data. 
K.B. and M.R. thank Franco Bonaf\'e and Orlando J. Silveira for insightful discussions and feedback on the developments and results we show in this work. 
Y.L. has been partly funded by the Deutsche Forschungsgemeinschaft (DFG, German Research Foundation) project number 467724959.

\end{acknowledgments}

\section*{References}

\end{bibunit}

\clearpage

\setcounter{section}{0}
\setcounter{equation}{0}
\setcounter{figure}{0}
\setcounter{table}{0}
\setcounter{page}{1}

\renewcommand{\thesection}{S\arabic{section}}
\renewcommand{\theequation}{S\arabic{equation}}
\renewcommand{\thefigure}{S\arabic{figure}}
\renewcommand{\thepage}{S\arabic{page}}
\renewcommand{\citenumfont}[1]{S#1}
\renewcommand{\bibnumfmt}[1]{$^{\rm{S#1}}$}

\title{Supporting Information for: \mytitle}
{\maketitle}

\begin{bibunit}

\section{Theoretical background}
\label{si-sec:theory}

In the following paragraph, we rewrite the derivation  of Equation~\ref{eq:hamiltonian} presented in our previous work~\cite{Litman2023/10.1021/acs.jpclett.3c01216}, in order to achieve a self-contained explanation of this implementation.
Our approach starts with the full, interacting and time-dependent problem described by the dipole-approximation Hamiltonian
\begin{equation}
    \hat{H}(t)
    =
    \hat{H}_0^{\mathrm{sc}}
    +
    \hat{H}_0^{\mathrm{tip}}
    -
    E_z \cos(\omega_\mathrm{p}t) \left( \hat{\mu}_z^{\mathrm{sc}} + \hat{\mu}_z^{\mathrm{tip}} \right)
    + \hat{V}(t),
    \label{si-eq:main}
\end{equation}
where $\hat{H}_0$ are the Hamiltonians of the unperturbed tip and the scattering (sc) subsystem, $E_z$ is the magnitude of the electric far field component of the incoming laser radiation oscillating at the plasmonic frequency $\omega_\mathrm{p}$, $\hat{\mu}_z$ are the dipole operators of the two subsystems and $\hat{V}(t)$ is the interaction between them.
Note that we limit ourselves to the $z$-components of all vector quantities: these represent the direction of the surface normal and are experimentally most relevant ones as TERS measurements are normally realized in the vertical back-scattering regime.
Next, we reduce the above expression into a much simpler static perturbation problem through several reasonable assumptions.
First, we assume a long tip-molecule distance which allows us to express the interaction only as the electrostatic (\textit{i.e.}, Hartree) potential of the tip under the influence of $E_z \cos(\omega_\mathrm{p}t)$ in the position representation as
\begin{equation}
    V(\mathbf{r}, t; \mathbf{R_\mathrm{tip}})
    \approx
    \Phi(\mathbf{r}, t; \mathbf{R}_\mathrm{tip})
    \equiv
    \frac{1}{4\pi\varepsilon_0} \int \mathrm{d}\mathbf{r}' \ \frac{\rho(\mathbf{r}, t; \mathbf{R}_\mathrm{tip})}{| \mathbf{r} - \mathbf{r}'|}.
\end{equation}
This is possible, because the tip density and the scattering-subsystem density do not overlap at large separations.
Next, we assume that the incoming radiation is weak enough as to ensure the linear polarization regime.
With that, we can Taylor-expand the field dependence of $\Phi$ to the first order around zero as follows
\begin{equation}
     \Phi(\mathbf{r}, t; \mathbf{R}_\mathrm{tip})
     \approx
     \Phi_0(\mathbf{r}; \mathbf{R}_\mathrm{tip})
     +
     E_z \left[ \frac{\partial \Phi(\mathbf{r}, t; \mathbf{R}_\mathrm{tip})}{\partial E_z} \right]_{E_z = 0}.
     \label{si-eq:taylor}
\end{equation}
In addition, we assume that the time evolution of $\Phi$ is dominated by a single frequency, \textit{i.e.}, that its Fourier transform can be well approximated as a Dirac $\delta$:
\begin{equation}
\begin{split}
    &\left[ \frac{\partial \tilde{\Phi}(\mathbf{r}, \omega; \mathbf{R}_\mathrm{tip})}{\partial E_z} \right]_{E_z = 0} \\
    &=
    \int_{-\infty}^{+\infty} \mathrm{d}t \ e^{-\mathrm{i} \omega t} \left[ \frac{\partial \Phi(\mathbf{r}, t; \mathbf{R}_\mathrm{tip})}{\partial E_z} \right]_{E_z = 0} \\
    &\approx
    \left[ \frac{\partial \tilde{\Phi}(\mathbf{r}, \omega_\mathrm{p}; \mathbf{R}_\mathrm{tip})}{\partial E_z} \right]_{E_z = 0} \delta(\omega - \omega_\mathrm{p}),
\end{split} 
\end{equation}
which is reasonable in materials with a strong and isolated plasmonic response such as Ag. 
In turn, this allows us to simplify the time-dependence in Equation~\ref{si-eq:taylor} to
\begin{equation}
\begin{split}
     \Phi(\mathbf{r}, t; \mathbf{R}_\mathrm{tip})
     &\approx
     \Phi_0(\mathbf{r}; \mathbf{R}_\mathrm{tip}) \\
     &+
     E_z \cos(\omega_\mathrm{p} t) \left[ \frac{\partial \tilde{\Phi}(\mathbf{r}, \omega_\mathrm{p}; \mathbf{R}_\mathrm{tip})}{\partial E_z} \right]_{E_z = 0}.
\end{split}
\end{equation}
These transformations allow us to split Equation~\ref{si-eq:main} into separate expressions for the isolated tip and the scattering subsystem under the influence of an external potential.
The time-dependent tip problem is solved using real-time TDDFT on the side and the key quantity $\tilde{\Phi}$ is numerically stored: the data format of our choice is a Gaussian Cube file.
The Equation for the scattering subsystem then reads
\begin{equation}
\begin{split}
    H^\mathrm{(sc)}(t)
    &=
    H^\mathrm{(sc)}_0
    +
    \Phi_0(\mathbf{r}; \mathbf{R}_\mathrm{tip}) \\
    &+
    E_z \cos(\omega_\mathrm{p} t) \left\{ 
    -\mu_z^\mathrm{(sc)} + \left[ 
    \frac{\partial \tilde{\Phi}(\mathbf{r}, \omega_\mathrm{p}; \mathbf{R}_\mathrm{tip})}{\partial E_z}
    \right]_{E_z=0}
    \right\}.
\end{split}
\end{equation}
The lack of hats over the operators in this expression implies assuming the position representation for all terms.
Finally, we restrict ourselves to non-resonant Raman scattering which allows us to formulate the corresponding time-independent problem
\begin{equation}
\begin{split}
    H^\mathrm{(sc)}
    &=
    H^\mathrm{(sc)}_0
    +
    \Phi_0(\mathbf{r}; \mathbf{R}_\mathrm{tip}) \\
    &+
    E_z \left\{ 
    -\mu_z^\mathrm{(sc)} + \left[ 
    \frac{\partial \tilde{\Phi}(\mathbf{r}, \omega_\mathrm{p}; \mathbf{R}_\mathrm{tip})}{\partial E_z}
    \right]_{E_z=0}
    \right\}
\end{split}
\end{equation}
which is directly addressable by static DFT and was implemented in FHI-aims~\cite{abbott2025roadmap-b1e} with a Gaussian cube file input for the spatially varying components of the external tip potential for the work in Reference~\cite{Litman2023/10.1021/acs.jpclett.3c01216}. 
That implementation only worked for open boundary conditions. 
In this contribution, we have instead derived and implemented a finite-field version for the calculation of the perturbation term and added it to the infrastructure in FHI-aims.
The calculation of the corresponding relevant polarizability components and the subsequent TERS intensity is described in detail in the main text.

\section{Computational workflow}
\label{si-sec:methodology}

In this section, we provide a comprehensive overview of the computational details, including geometry relaxation, vibrational analysis and the TERS imaging simulations.

\subsection{Geometry preparation and Hessian calculation}

All geometries were combined using tools provided by the ASE Python package~\cite{Larsen2017/10.1088/1361-648X/AA680E} and optimized within the FHI-aims software~\cite{Blum2009/10.1016/j.cpc.2009.06.022}.
In order to correctly apply the Born--von-Kármán periodic boundary conditions, the surface systems need to be large enough to accommodate the numerically tabulated tip potential within a single unit cell.
For the Ag(100) slab used in the simulations of TCNE and MgP, this means a size of 4 vertically organized layers of 8$\times$8 atoms, leading to the total of 256 Ag atoms inside a periodic cell with a horizontal side length of 23.51~\AA.
For the MoS$_2$ systems, we used 15$\times$15 unit cells with a total of 675 atoms, a lateral side length of 47.7~\AA\ and the angle between the lateral unit cell vectors of 120$^\circ$.
A 50~\AA -thick vacuum layer was added to each side of all slabs.

Regarding the electronic structure, we employed the \textit{``light''} settings and the PBE generalized-gradient-approximation density functional~\cite{Perdew1996/10.1103/PhysRevLett.77.3865} equipped with the Tkatchenko--Scheffler dispersion correction~\cite{Tkatchenko2009/10.1103/PHYSREVLETT.102.073005} for the TCNE and MgP systems.
This is a reasonably standard and well-tested set-up for surface-bound adsorbates which allows us for a direct comparison with our previous work where it was also employed~\cite{Litman2023/10.1021/acs.jpclett.3c01216}.
In addition, we found that for the MoS$_2$ slabs, especially those including vacancies, a much more converged electronic structure set up was needed to obtain well-behaved vibrational properties.
For that reason, we used the \textit{``tight''} basis set, increased the \texttt{radial\_multiplier} keyword~\cite{Blum2009/10.1016/j.cpc.2009.06.022} to the value of 4 and the \texttt{wave\_threshold} to $10^{-9}$ au.
To ensure tight convergence of the electronic density, we employed a self-consistent cycle convergence parameter of 10$^{-7}$~$e a_0^{-3}$ in all systems.
The gas-phase systems were treated under open boundary conditions.
The slabs were treated with a full 3D periodicity and a dipole correction~\cite{Neugebauer1992/10.1103/PhysRevB.46.16067} was employed across the vacuum-containing direction (\textit{i.e.}, $z$) to prevent spurious interactions between neighboring replicas. 
We stress that the dipole correction is necessary to yield the correct response behavior of the electronic structure of the systems under small applied electric fields. 
For the required slab sizes, we found it was sufficient to consider only $\Gamma$-point Kohn--Sham states~\cite{Kohn1965/10.1103/PhysRev.140.A1133, Blum2009/10.1016/j.cpc.2009.06.022}.

We found that converging the symmetries of the TERS images, especially the ones with low intensities, requires a very tight geometry optimization.
For this reason, we employed a strict convergence criterion of 10$^{-4}$~eV$\mathrm{\AA}^{-1}$ and relied on the Broyden--Fletcher--Goldfarb--Shanno (BFGS) minimizer~\cite{Broyden1970/10.1093/IMAMAT/6.1.76, Fletcher1970/10.1093/COMJNL/13.3.317, Goldfarb1970/10.2307/2004873, Shanno1970/10.2307/2004840} to relax our structures.
The lengths of the lattice vectors of the Ag(100) slabs were first optimized without the TCNE and MgP adsorbates while keeping the angles fixed.
Then, we deployed the molecules on the Ag surfaces and continued the optimization with a fixed unit-cell size to fully optimize the adsorbed configurations.
For the MoS$_2$ slabs, we first optimized smaller slabs and later generalized to the aforementioned bigger systems for the TERS calculations.
For the pristine monolayer, we optimized a 5$\times$5 surface including the lattice vectors with fixed angles using a 3$\times$3$\times$1 $\Gamma$-centered k-point grid.
For the defective system carrying a sulfur monovacancy, we continued the optimization in a fixed unit cell after removing a single sulfur atom from the optimized pristine monolayer.

Once we had access to relaxed geometries, we proceeded to the calculation of the Hessian matrices of the studied systems.
To achieve this, we relied on the \texttt{Vibrations} module of ASE to generate Cartesian displacements and compute their energies and forces using an FHI-aims client with settings consistent with the ones described above.
In all Hessian calculations, we employed Cartesian displacements of 2$\cdot$10$^{-3}$~\AA\ for each degree of freedom.
To save computational time, we restricted the Hessian calculations in the case of the TCNE and MgP adsorbates on Ag(100) to the subspace of the configurational space that belong to the molecule.
In those cases, the Hessian is well approximated by a block-diagonal structure with minimal vibrational couplings between the surface and the molecule.
Therefore, the explicit surface must not necessarily \textit{move} in the Hessian calculation, but it must still be \textit{present} as it strongly affects the electronic structure and the geometry of the molecule.
For the pristine and defective MoS$_2$ slabs, we calculated the $\Gamma$-point modes of a smaller 5$\times$5 unit cell and extended the obtained A$_\mathrm{1g}$ mode (or the corresponding defective mode) to the full 15$\times$15 system needed for the TERS calculation.
In the case of the defective monolayer, this corresponds to a vacancy concentration of 4\%.

\subsection{Calculation of local polarizabilities}

To accomplish the calculation of the TERS images presented in the main text, we rely exclusively on our FHI-aims implementation of the finite-field version of the electronic calculation in the presence of plasmonic near fields as described in Section~\ref{si-sec:theory}.
For practical reasons, the $\Phi_0$ can be neglected to a good approximation (as we show in Section~\ref{si-sec:groundstate}).
Our implementation expects the volumetric tip potentials tabulated in a Gaussian cube format and our calculations rely on the exact same tip potential as used in our previous work~\cite{Litman2023/10.1021/acs.jpclett.3c01216} which was calculated using the Octopus software~\cite{Tancogne-Dejean2020/10.1063/1.5142502/954926}. 
No new time-dependent DFT calculations were performed in this work.

We have implemented an FHI-aims workflow for the reading, interpolation and relative positioning of these cube files with respect to the scattering subsystem.
Essentially, this means that a single FHI-aims control file is able to set up a full single-point electronic structure calculation including the tip potential that yields a value of $\mu_z(\mathbf{R}_\mathrm{tip})$ for a given tip position $\mathbf{R}_\mathrm{tip}$ (the tip position is usually defined by the tip apex, but note that no explicit atoms of the tip, rather just the electrostatic potential $\Phi$, enter the calculation).
Then, of course, scanning over a grid of values of $\mathbf{R}_\mathrm{tip}$ at two different values of $E_z$ allows to use the finite-field formulation of local polarizabilities (Equation~\ref{eq:polarizability} of the main text) and Raman intensities (Equation~\ref{eq:raman} of the main text) allows to construct a TERS image.
For this purpose, we employed in all systems a Cartesian displacement of the structures along normal mode vectors of length of 5$\cdot$10$^{-3}$~\AA\ and calculated the dipole moment $z$-components at field values of $0$ and 10$^{-1}$~V \AA$^{-1}$ for the finite difference evaluation.
For the TCNE in the gas phase and on Ag(100), we used a 20$\times$20 pixel grid spanning from $-$5.0 to 5.0~\AA\ in both lateral directions, for the MgP we used the same number of pixels, but spanning a larger area of $-$6.5 to 6.5~\AA\ and for the MoS$_2$ slabs, we used 12$\times$12 grid spanning $-$8.75 to 8.75~\AA\ in the $x$-direction and $-$7.57 to 7.57~\AA\ in the $y$-direction (\textit{cf.} Figure~\ref{fig:MoS2} for the corresponding monolayer orientation).
For the gas-phase calculations of TCNE and MgP, we always use the molecule in the surface-adsorbed geometry to meaningfully quantify the effect of the surface.

\section{Neglect of ground-state Hartree potential term}
\label{si-sec:groundstate}

\begin{figure}[tb!]
    \centering
    \includegraphics[width=\linewidth]{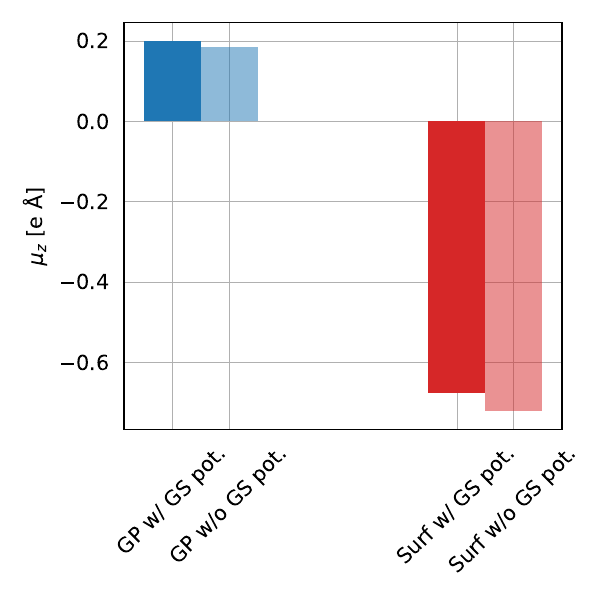}
    \caption{
    The dipole moment of the TCNE molecule in the gas phase (GP) and on the Ag(100) surface calculated with or without the ground state tip potential (\textit{cf.} second term of the right-hand side of Equation~\ref{eq:hamiltonian} of the main text).
    In all cases the magnitude of the applied homogeneous far field was set to 10$^{-3}$ V \AA$^{-1}$ and the derivative of the tip potential with respect to the far field was used as usual.
    }
    \label{si-fig:groundstate}
\end{figure}

For practical reasons, the TERS images in the main text were calculated while neglecting the $\Phi_0(\mathbf{r}; \mathbf{R}_\mathrm{tip})$ term in Equation~\ref{eq:hamiltonian} of the main text.
The reason for this is that at tip--molecule distances relevant to our calculations, the ground state potential only has a minor overlap with the electron density of the scattering subsystem.
On the other hand, the derivative of the potential with respect to the electric field extends spatially far below the tip apex and overlaps strongly with the electron density.
In Figure~\ref{si-fig:groundstate}, we show the dipole moment $z$-component values calculated for the TCNE molecule in the gas phase and on Ag(100) with the tip located 4~\AA\ above the center of the central C--C bond where the tip is at its closest.
The calculated value of the dipole moment only changes by 5--7\% after including the ground state term.
This implies that its contribution is more than an order of magnitude smaller that that of the potential derivative and neglecting it only introduces a minor effect, if any, on the resulting shape of the TERS images.

Note that the neglect of the ground state potential has a significant effect on the computational demands of our calculations.
Once neglected, the $E_z = 0$ part of the calculation loses all its dependence on $\mathbf{R}_\mathrm{tip}$.
In turn, this means that the zero-field part of the calculation is essentially only two single-point dipole evaluations at the two displaced geometries. 
Therefore, for a TERS image with a grid of $N^2$ pixels, while the full calculation requires the total of $4N^2$ single-point calculations ($N^2$ for two displacements and two field strengths), the calculation neglecting the ground state only requires $2N^2 +2$ single-points.
Given the required size of the surfaces needed for these calculations, this can represent a significant speed up.

\section{Validity of the linear polarization regime}

\begin{figure}[tb!]
    \centering
    \includegraphics[width=\linewidth]{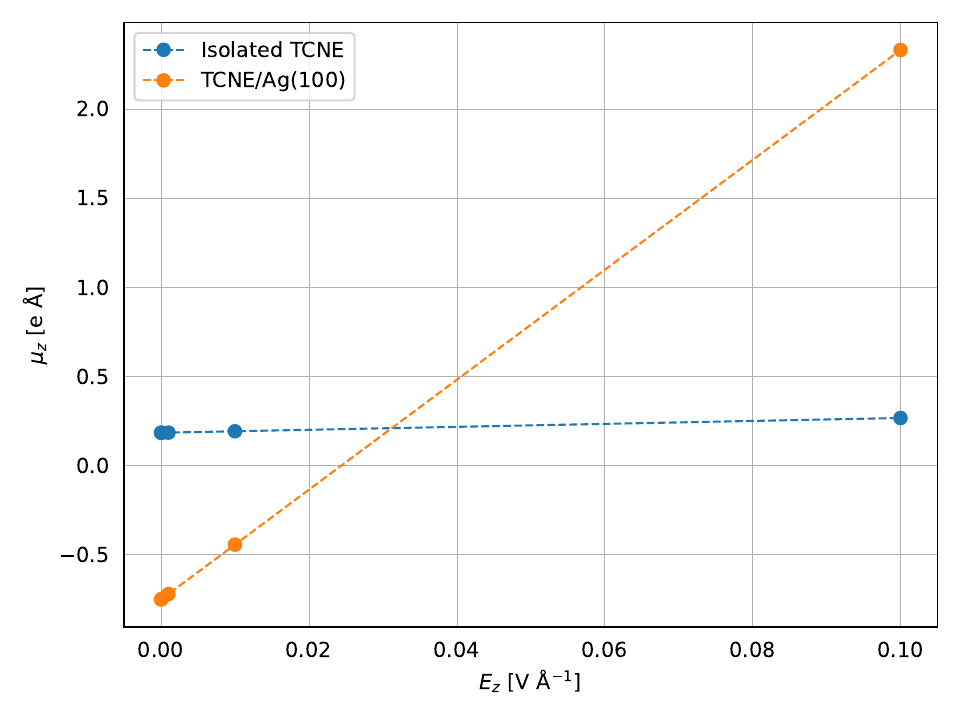}
    \caption{
    The dipole moment of the isolated TCNE molecule (blue) and Ag(100)-bound TCNE (orange) in the presence of the tip near field at applied homogeneous electric fields ranging from 0 to 0.1 V \AA$^{-1}$.
    }
    \label{si-fig:dipole-linearity}
\end{figure}

Another approximation behind Equation~\ref{eq:hamiltonian} of the main text is that the applied far fields are weak enough that the system polarizes linearly.
This allows us to use the first-order Taylor expansion in Equation~\ref{si-eq:taylor}.
In Figure~\ref{si-fig:dipole-linearity}, we show the dependence of $\mu_z$ on the magnitude of the applied electric far field for the TCNE both in the gas phase and on Ag(100).
Once again, the tip (now only represented by the potential derivative with the ground state contribution neglected) is located 4~\AA\ above the central C--C bond.
For both systems, we are safely in the linear polarization regime up to fields of 10$^{-1}$ V \AA$^{-1}$.

\section{Correspondence of DFPT and finite-field calculations for gas-phase TCNE}
\label{si-sec:dfpt-vs-ff}

\begin{figure}[tb!]
    \centering
    \includegraphics[width=\linewidth]{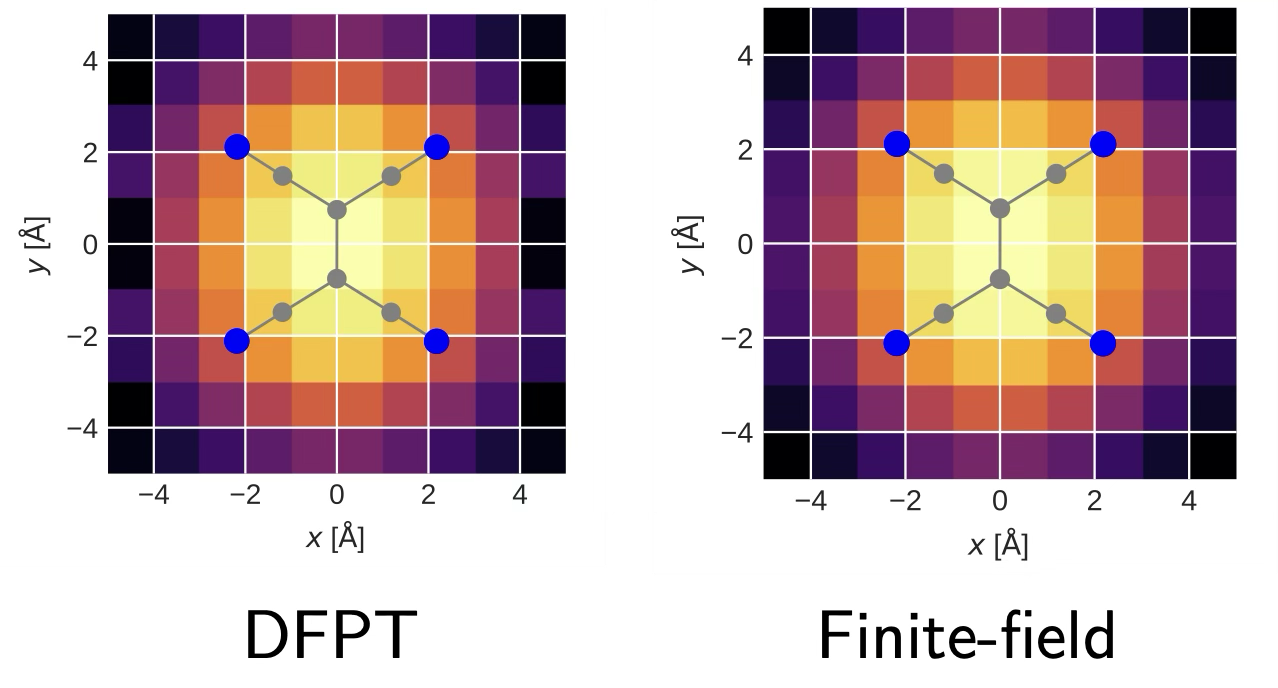}
    \caption{
    The A$_1$ mode TERS image of gas-phase TCNE calculated using DFPT (left) and the FF-based approach (right).
    The latter is equivalent to Figure~\ref{fig:TCNE}C of the main text.
    The intensities are encoded through the color map, increase towards lighter colors and have the same range.
    }
    \label{si-fig:dfpt-vs-ff}
\end{figure}

Here, we present a comparison of the present implementation using FFs and the previous one using DFPT in Figure~\ref{si-fig:dfpt-vs-ff}.
Both methods aim to calculate the $zz$-component of the polarizability tensor and are, therefore, equivalent.
We show this on the case of the A$_1$ mode of gas-phase TCNE, where we employ the FF-based and the DFPT-based approach to the exactly same geometry and find that, within numerical accuracy given by the different nature of the two calculations, the obtained polarizabilities are indeed identical.
However, we note again that only the FF-TERS approach can be employed to periodic systems.

\section{DFPT-based TERS images of TCNE on silver clusters}
\label{si-sec:old-ters}

\begin{figure*}
    \centering
    \includegraphics[width=\linewidth]{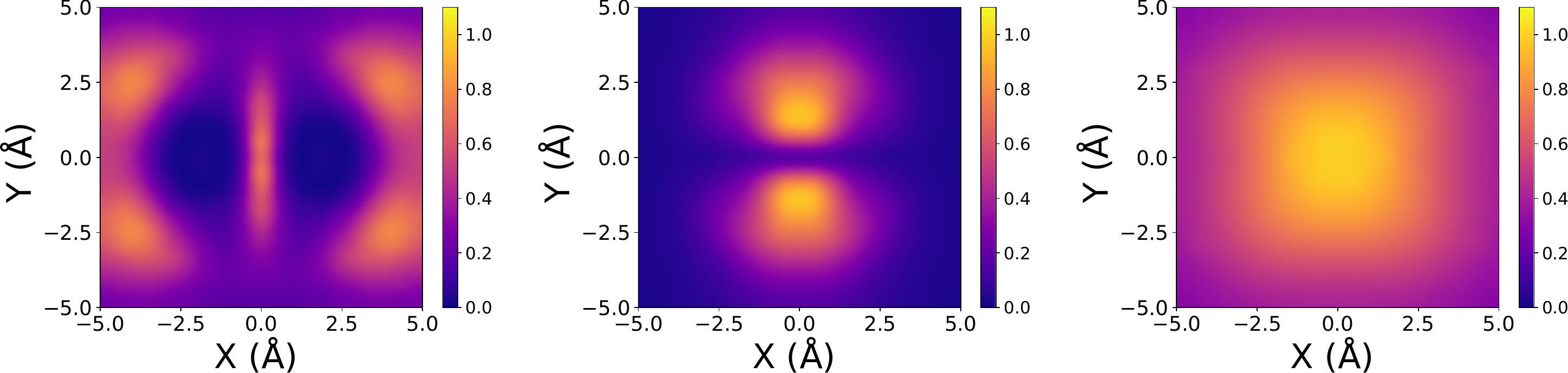}
    \caption{
    TERS images of TCNE on a 4$\times$4$\times$3 Ag(100) cluster corresponding one to one to the periodic Ag(100) surface images shown in Figure~\ref{fig:TCNE} in the main text.
    The intensities in this plot are normalized to the maximum intensity in each image.
    Adapted from Reference~\citenum{Litman2023/10.1021/acs.jpclett.3c01216}.
    }
    \label{si-fig:original-tcne}
\end{figure*}

In Figure~\ref{si-fig:original-tcne}, we plot the original TERS images of the same vibrational modes of TCNE as discussed in the main text on an Ag cluster calculated using the DFPT-based workflow in open boundary conditions.
Note the dramatic change in the leftmost panel in comparison to the B$_1$ data obtained on the periodic surface in Figure~\ref{fig:TCNE}.

\section{Pristine molybdenum disulfide monolayer TERS imaging}
\label{si-sec:pristine-mos2}

\begin{figure}[tb!]
    \centering
    \includegraphics[width=0.75\linewidth]{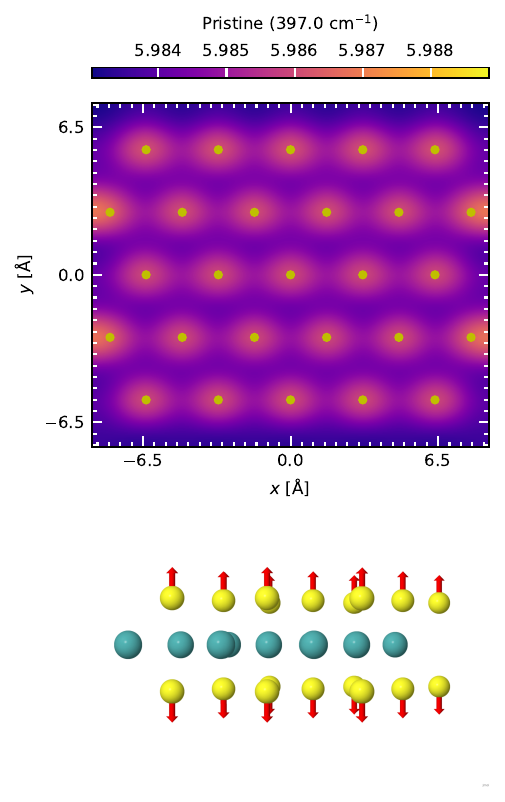}
    \caption{ 
        TERS imaging of the Raman-active A$\mathrm{1g}$ vibration in a pristine MoS$_2$ monolayer.
        The top panel shows the calculated TERS image. 
        The positions of the top-layer sulfur atoms are given by the yellow circles.
        The TERS intensity is shown in the units of 10 $\mathrm{e}^2 \mathrm{\AA}^2 \mathrm{V}^{-2}$.
        The bottom panel shows a side view the vibration on a 3$\times$3 unit cell.
        In these snapshots, the sulfur atoms are shown in yellow, the molybdenum atoms in turquoise and the cartesian atomic components of the normal mode vector as red arrows.
    }
    \label{si-fig:MoS2}
\end{figure}

In Figure~\ref{si-fig:MoS2}, we present the complementary TERS image of the pristine A$_\mathrm{1g}$ vibration in MoS$_2$.

\section{Image asymmetry after using a realistic near field vs. a symmetric dipolar approximation}
\label{si-sec:tip-symmetry}

\begin{figure}[tb!]
    \centering
    \includegraphics[width=0.75\linewidth]{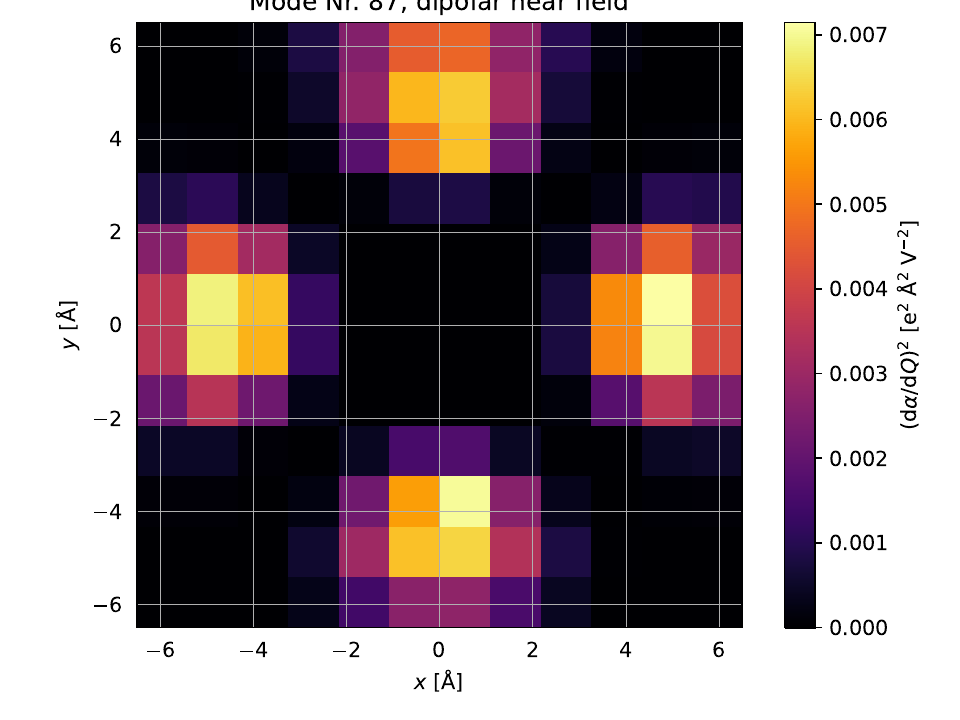}
    \caption{
    B$_\mathrm{1g}$ mode MgP/Ag(100) TERS image calculated with a dipolar field approximating the realistic field used for the production of the results shown in the main text.
    }
    \label{si-fig:dipolar}
\end{figure}

We have calculated the TERS image of the B$_\mathrm{1g}$ mode of MgP/Ag(100) calculated with the $C_\infty$-symmetric dipolar-field model of the tip potential that does not break any underlying molecular symmetries.
As shown in Figure~\ref{si-fig:dipolar}, using such a near field indeed leads to a fully symmetric image in which the four peaks are the same shape and at equivalent positions in comparison to the one obtained with a realistic near field and shown in Figure~\ref{fig:MgP}E of the main text.
The remaining (minute) imperfections in the image we understand as a testament to the sensitivity of the FF-TERS calculations to residual strain from the finite-threshold numerical optimization of the molecular geometry, imprecision in the calculation of normal modes and numerical noise in the integration of the electron density to obtain $\mu_z$.

\section{Dependence of the shape of TERS images on the equilibrium binding distance from the surface}
\label{si-sec:distance}

\begin{figure}[tb!]
    \centering
    \includegraphics[width=0.75\linewidth]{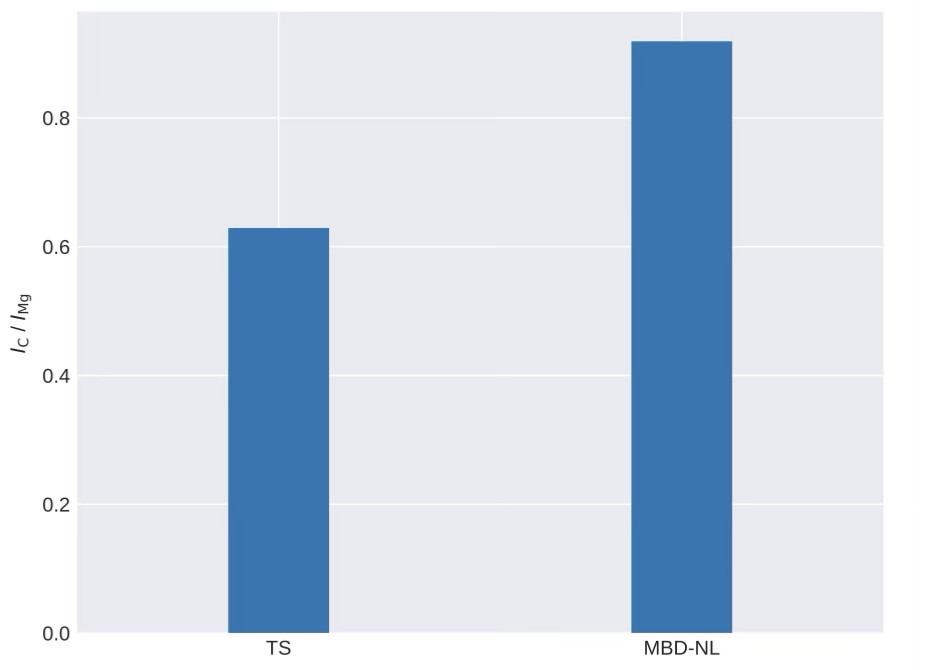}
    \caption{
    Dependence on the intensity pattern of the A$_\mathrm{2u}$ of MgP/Ag(100) on the surface--molecule binding distance: the ratio of TERS intensities above the bridge carbon atom and at the origin (above the Mg atom) as calculated by PBE-TS and PBE-MBD.
    }
    \label{si-fig:distance}
\end{figure}

\begin{table}[tb!]
    \centering
    \begin{tabular}{cc}
        \toprule
        Dispersion correction & Equilibrium distance [\AA] \\
        \midrule
         Hirshfeld TS~\cite{Tkatchenko2009/10.1103/PHYSREVLETT.102.073005} & 2.77 \\
        \midrule
        Non-local MBD~\cite{Tkatchenko2012/10.1103/physrevlett.108.236402} & 2.89 \\
        \bottomrule
    \end{tabular}
    \caption{
    Equilibrium surface--molecule binding distances in MgP/Ag(100) for different dispersion corrections.
    }
    \label{si-tab:distances}
\end{table}

Here, we further support the discussion about the dependence the TERS images are on the chosen level of electronic structure that was outlined in Figure~\ref{fig:decomposition}A) of the main text.
Here, we show an additional result that corroborates the observation that the experimentally observed tails to the binding distance between the MgP molecule and the surface.
Specifically, in Figure~\ref{si-fig:distance}, we show that with a change of the dispersion correction from Tkatchenko--Scheffler~\cite{Tkatchenko2009/10.1103/PHYSREVLETT.102.073005} (TS) to many-body dispersion~\cite{Tkatchenko2012/10.1103/physrevlett.108.236402} --- which generally tends to increase non-covalent binding distances (see Table~\ref{si-tab:distances}) --- the image gains more intensity over the bridge carbons as the ratio of the TERS intensities $I_\mathrm{C} / I_\mathrm{Mg}$ grows from $\sim$0.6 to $\sim$0.9.

\section{Changes in electron densities in Mg porphine}
\label{si-sec:density-changes}

\begin{figure}[tb!]
    \centering
    \includegraphics[width=\linewidth]{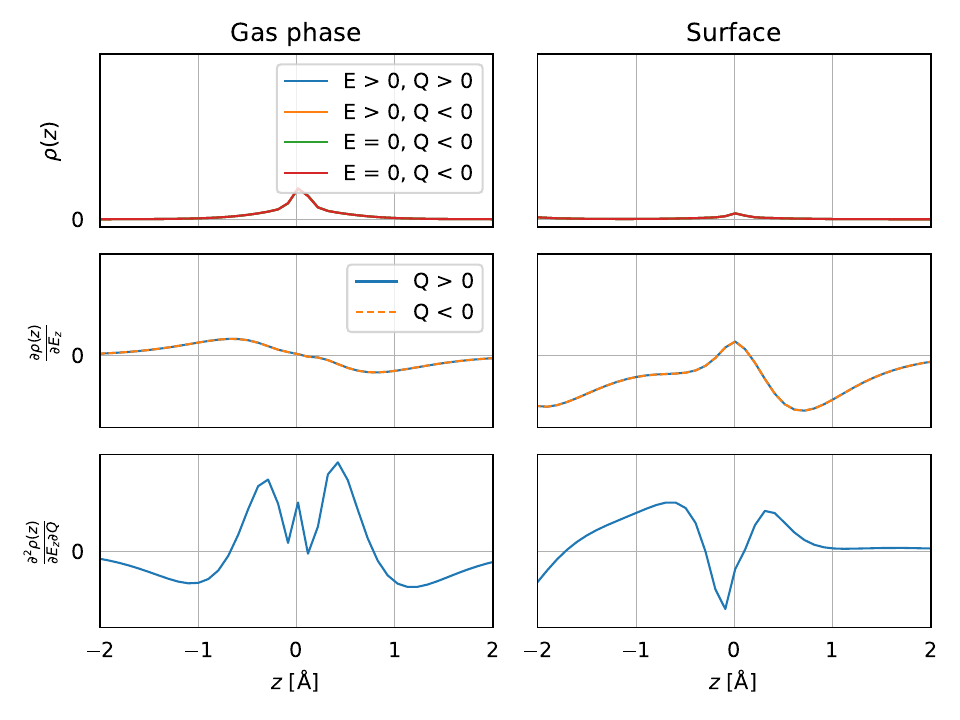}
    \includegraphics[width=\linewidth]{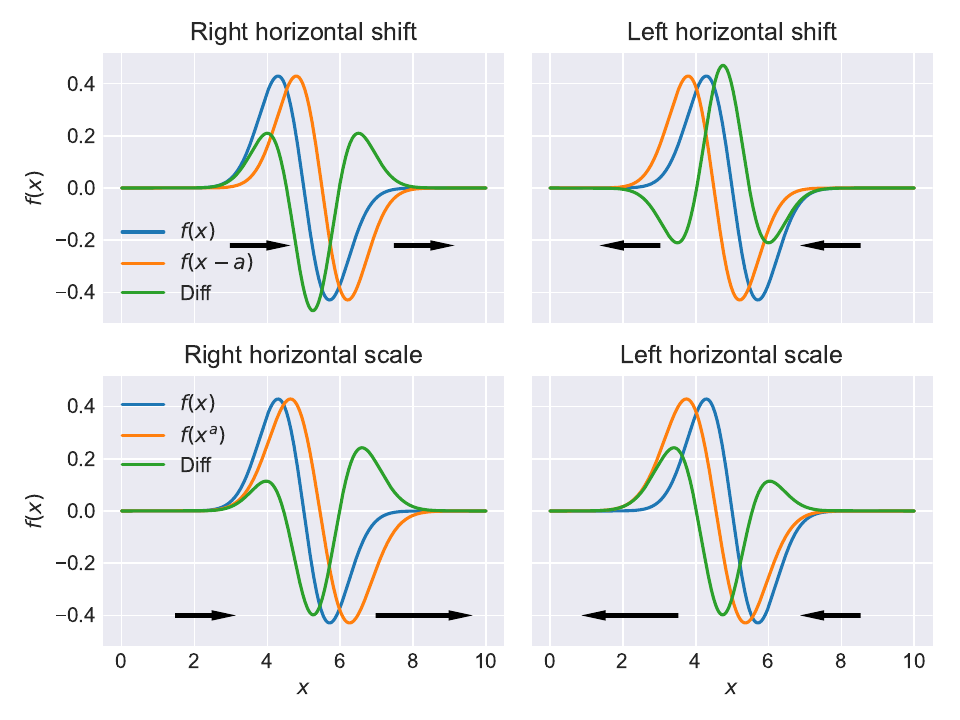}
    \caption{
    Top 3$\times$2 grid: the plots of the electron density $\rho$ as a function of the vertical position $z$, of the density response $\partial \rho / \partial E_z$ and the change of the density response with the molecular vibration $\partial^2 \rho / \partial E_z \partial Q_k$.
    Bottom 2$\times$2 panel: a simple model showing the origin of the $\partial^2 \rho / \partial E_z \partial Q_k$ curve shapes. The orange and blue curves represent the function of the general shape $f(x) = -x e^{-x^2}$ with various shifting or scaling transformations; the green curves always shows their difference.
    The top two panels show that a shift can give rise to the general two-peak pattern, but cannot alone give rise to unevenness in it.
    The bottom panels show that a horizontal scaling is needed for the unevenness to kick in: this corresponds to a larger electronic polarizability in the realistic case.
    }
    \label{si-fig:density-changes}
\end{figure}

In Figure~\ref{si-fig:density-changes}, we demonstrate the origin of the sign change in the quantity $A_{zz}$ discussed in the main text.
In the first row, we show the electron density as a function of the $z$-coordinate
\begin{equation}
    \rho(z)
    \equiv
    \int_{-\infty}^{\infty} \int_{-\infty}^{\infty} \dd x \dd y \ \rho(x, y, z)
\end{equation}
with the molecule centered at $z=0$
for the case with and without the homogeneous far field $E_z$ and at both positive and negative displacements of the normal coordinate $Q$.
In the second row, we show the density response
\begin{equation*}
    \frac{\partial \rho(z)}{\partial E_z}
\end{equation*}
due to the action of the far field $\mathbf{E} = (0, 0, E_z)$.
In the gas phase, this features a clear ``beat'' pattern corresponding to the polarization of the molecule and the concentration of negative charge in the $z < 0$ region and of positive charge at the $z > 0$ region.
Almost identical pattern is obtained at both displacements $\pm Q$.
This observation is qualitatively identical on the surface; however, the negative cloud effectively spills over into the metal, leading to an uneven beat with a more pronounced $z > 0$ side.
Finally, in the third row, we plot the second mixed derivative of the density
\begin{equation*}
    \frac{\partial^2 \rho(z)}{\partial E_z \partial Q_k}.
\end{equation*}
This quantity integrates over $z$ to give $A_{zz}$ and features a two-peak pattern that arises as a horizontal scaling possibly also accompanied by a horizontal shift of the beats at various displacements.
In the gas phase, this two-peak pattern in larger at the $z > 0$ side, indicating that the geometry corresponding to $Q > 0$ permits a further displacement of electronic density along the positive direction of $z$: in other words, the $Q > 0$ is more polarizable in the gas phase.
Contrarily, on the surface, the pattern is larger at the $z < 0$ side.
This suggests that the $Q > 0$ geometry has its electron density pulled stronger towards the polarized slab that thus screens the change in the density induced by the far field.
All in all, this leads to an overall smaller induced dipole in the $Q > 0$ case and, consequently, smaller polarizability and the observed negative $A_{zz}$.

The bottom panel illustrates the origin of the two-peak patterns observed in the second mixed density derivatives using a simple model for the beat of the form of
\begin{equation}
    f(x)
    =
    -x e^{-x^2}
\end{equation}
centered around $x=5$.
In the first row we show that a simple shift $x \rightarrow x - a$ indeed gives rise to the two-peak pattern, but a one that is strictly even, \textit{i.e.}, with both sides of the same height.
In order to explain the observed unevenness in the realistic case, one must consider a horizontal scaling $x \rightarrow x^{-a}$ that makes the sides uneven.
Clearly, this corresponds to the larger or smaller displacements of electron density and the observed differences in polarizabilities.
Finally, we note that the region around $z=0$ does not behave in full accordance with the simple models.
This is due to the behavior of the core electrons of the molecule localized in closed proximity of the atoms and is not relevant for the present discussion.

\section{Atomic decomposition of TERS amplitude}
\label{si-sec:atomic-decomposition}

\begin{figure}[tb!]
    \centering
    \includegraphics[width=\linewidth]{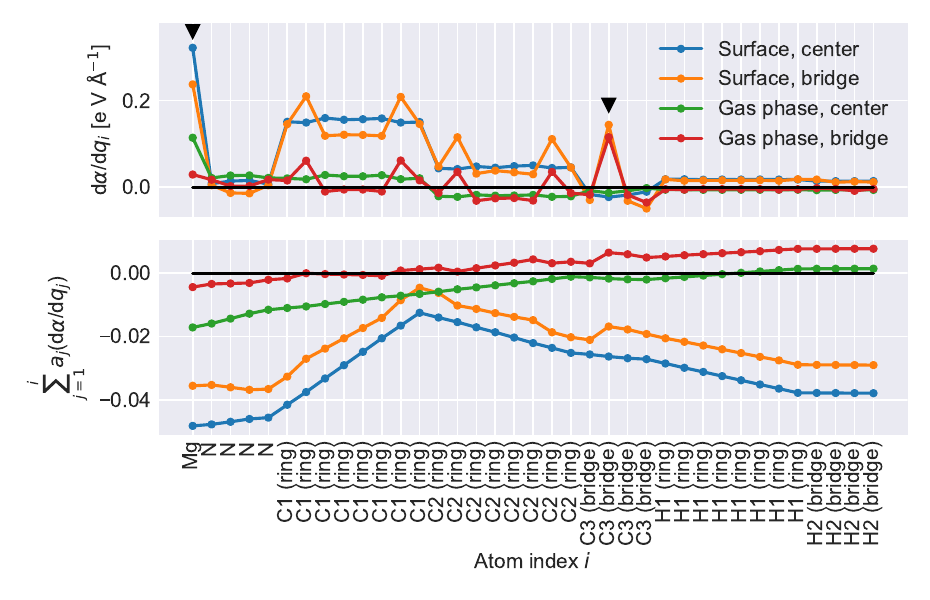}
    \caption{
    Atomic decomposition of the TERS amplitude.
    Top panel: magnitudes of the individual atomic terms.
    Bottom panel: cumulative sum to show how the overall amplitude for the given tip position is built up from the individual terms.
    Note that C1 is the pyrrole carbon closer to the center of the molecule, the rest of the labels is self explanatory.
    }
    \label{si-fig:atomic-amplitudes}
\end{figure}

In the top panel of Figure~\ref{si-fig:atomic-amplitudes}, we show in the individual atomic terms 
\begin{equation*}
    \frac{\partial \alpha_{zz}^\mathrm{local}(\mathbf{R}_\mathrm{tip})}{\partial z_i},
\end{equation*}
of the decomposition of the A$_\mathrm{2u}$ amplitude of MgP in the gas phase and on Ag(100) for two different tip positions.
In the above expression, $i$ indexes the atoms in the molecule: note that only the $z$-direction is relevant for this normal mode.
We note that the adsorption makes the terms corresponding to the C2, H1 and H2 atoms change signs and those atoms that are in the closest neighborhood of $\mathbf{R}_\mathrm{tip}$ give a stronger response.
In the bottom panel, we plot the cumulative sum taken over the atomic index $i$ with each term multiplied by the normal mode displacement.
As such, at the end of each curve is the full value of $A_{zz}$ in accordance with Equation~\ref{eq:decomposition} in the main text.
Note how the sign-change effect is formed and the surface ends up negative while the gas phase ends up positive.

\section{Further interpretation of TERS cross terms}
\label{si-sec:locality-of-density-response}

\begin{figure}[tb!]
    \centering
    \includegraphics[width=\linewidth]{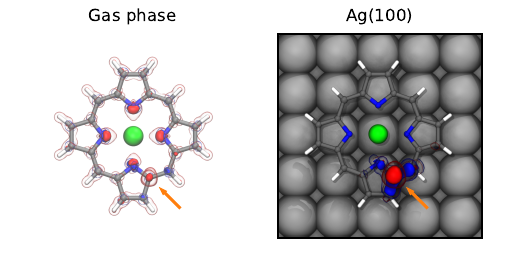}
    \caption{
    Spatial dependence of the quantity $\partial^2 \rho(\mathbf{r};\mathbf{R}_\mathrm{tip}) / \partial E_z \partial z_i$ where the index $i$ corresponds to the pyrrole (C1, see previous Section) carbon atom highlighted by the orange arrow (only the $z$-displacement is relevant for the A$_\mathrm{2u}$ mode, \textit{cf.} Figure~\ref{fig:MgP}J).
    In the gas phase, the shown contours are at $\pm 1$ and $\pm 3 \times 10^{-6}$ e \AA$^{-3}$ V$^{-1}$ (red for positive, blue for negative and with the higher isovalue as the solid surface), in the Ag(100)-bound system $\pm 1$ and $\pm 3 \times 10^{-5}$ e \AA$^{-3}$ V$^{-1}$.
    }
    \label{si-fig:density-changes2}
\end{figure}

In this Section, we provide an open insight into the balance between self and cross terms of the decomposition of the TERS intensity as discussed in the main text.
The question we are trying to answer here is what exactly the surface does that the cross term become smaller and less relevant in shaping the surface-bound image?
To provide, we note that
\begin{equation}
    \frac{\partial \alpha_{zz}^\mathrm{local}(\mathbf{R}_\mathrm{tip})}{\partial q_i}
    =
    \int_\text{unit cell} \dd \mathbf{r} \ z \frac{\partial^2 \rho(\mathbf{r}; \mathbf{R}_\mathrm{tip})}{\partial E_z \partial q_i}
\end{equation}
and inspect the second mixed derivative inside the integrand as a real-space function of $\mathbf{r}$.
This quantity (also encountered previously in a slightly different context in Section~\ref{si-sec:density-changes}) capture how the electronic density response to a far-field perturbation changes with the molecular vibration.
To illustrate its nature, we picked a single atom of the molecule (one of the C1 pyrrole carbons) and moved it from its equilibrium position by $\pm 5 \cdot10^{-3}$~\AA\ to simulate the variation of $q_i$ and calculate the numeric derivative; the volumetric plot is shown in Figure~\ref{si-fig:density-changes2} for both the gas phase and the surface.
While the quantity is fully delocalized over the entire molecule in the gas phase, on the surface we observe a localization to a small surrounding of the plucked atom.
This is observed for all atom types in this system.
This surface-induced localization could explain the loss of magnitude in the cross term. 
As shown in Section~\ref{si-sec:atomic-decomposition} above, the atomic terms corresponding to atoms located right under the tip apex get enhanced, suggesting that the self terms of atoms in proximity of the tip will be rather strong and the cross terms containing contribution from distant atoms rather weak.
However, there are many more cross terms than self terms, so the outcome will depend on a fine balance of the individual contributions.
Testing this hypothesis will require an extensive calculation of the individual terms as a function of $\mathbf{R}_\mathrm{tip}$, which was not done in this work.

\section*{References}

\end{bibunit}

\end{document}